\documentclass[aps,prx,twocolumn,footinbib,superscriptaddress,showkeys,longbibliography]{revtex4-1}

\usepackage{graphicx}
\usepackage{indentfirst}
\usepackage{braket}
\usepackage{float}
\usepackage{amsmath}
\usepackage{amssymb}
\usepackage{amsfonts}
\usepackage{epstopdf}
\usepackage{CJK}
\usepackage{esint}
\usepackage{color}
\usepackage[T1]{fontenc}
\usepackage{subfigure}
\usepackage{amsfonts}
\usepackage{natbib}
\usepackage{footmisc}
\usepackage{scrextend}
\usepackage{verbatim}
\usepackage{multirow}
\usepackage[english]{babel}
\usepackage{url}
\usepackage{bm}
\usepackage{hyperref}
\definecolor{darkblue}{rgb}{0,0,0.5}
\hypersetup{
colorlinks=true,
linkcolor=black,
filecolor=blue,
citecolor=darkblue,  
urlcolor=black,
}

\usepackage{enumitem,kantlipsum}
\usepackage{refcount}
\usepackage{mfirstuc}

\newenvironment{proof}[1][{\em Proof.---}]{{#1} }{\ }

\newcounter{lemmacounter}
\newenvironment{lemma}[1][]{\refstepcounter{lemmacounter}
   {{\em Lemma~\thelemmacounter #1}.---} \rmfamily}
   
\newcounter{theoremcounter}
\newenvironment{theorem}[1][]{\refstepcounter{theoremcounter}
   {{\em Theorem~\thetheoremcounter #1}.---} \rmfamily}
   
\newcounter{conjecturecounter}
\newcounter{definitioncounter}
\newenvironment{definition}[1][]{\refstepcounter{definitioncounter}
   {{\em Definition~\thedefinitioncounter #1}.---} \rmfamily}

\def\be{\begin{equation}}
\def\ee{\end{equation}}
\def\ba{\begin{eqnarray}}
\def\ea{\end{eqnarray}}
\newcommand{\calI}{{\cal I}}

\newcommand{\calT}{{\cal T}}
\newcommand{\calG}{{\cal G}}
\newcommand{\calH}{{\cal H}}
\newcommand{\tr}{{\rm Tr}}

\newcommand{\1}{^{(1)}}
\newcommand{\tdelta}{\tilde{\delta}}

\begin{document}
%\preprint{MIT-CTP/4835}
\date{\today}
\title{
Resource theory of non-Gaussian operations
}

\author{Quntao Zhuang}
\email{quntao@mit.edu}
\affiliation{
Department of Physics, 
Massachusetts Institute of Technology, Cambridge, MA 02139, USA}
\affiliation{Research Laboratory of Electronics, 
Massachusetts Institute of Technology, Cambridge, MA 02139, USA}

\author{Peter W. Shor}
\affiliation{Center For Theoretical Physics,
Massachusetts Institute of Technology, Cambridge, MA 02139, USA}
\affiliation{
Department of Mathematics, 
Massachusetts Institute of Technology, Cambridge, MA 02139, USA}

\author{Jeffrey H. Shapiro}
\affiliation{Research Laboratory of Electronics, 
Massachusetts Institute of Technology, Cambridge, MA 02139, USA}

\date{\today}

\begin{abstract}
Non-Gaussian states and operations are crucial for various continuous-variable quantum information processing tasks. To quantitatively understand non-Gaussianity beyond states, we establish a resource theory for non-Gaussian operations. In our framework, we consider Gaussian operations as free operations, and non-Gaussian operations as resources. We define entanglement-assisted non-Gaussianity generating power and show that it is a monotone that is non-increasing under the set of free super-operations, i.e., concatenation and tensoring with Gaussian channels. For conditional unitary maps, this monotone can be analytically calculated. As examples, we show that the non-Gaussianity of ideal photon-number subtraction and photon-number addition equal the non-Gaussianity of the single-photon Fock state. Based on our non-Gaussianity monotone, we divide non-Gaussian operations into two classes: (1) the finite non-Gaussianity class, e.g., photon-number subtraction, photon-number addition and all Gaussian-dilatable non-Gaussian channels; and (2) the diverging non-Gaussianity class, e.g., the binary phase-shift channel and the Kerr nonlinearity. This classification also implies that not all non-Gaussian channels are exactly Gaussian-dilatable. Our resource theory enables a quantitative characterization and a first classification of non-Gaussian operations, paving the way towards the full understanding of non-Gaussianity.

%\begin{description}
%\item[Subject Areas]
%Quantum Information, Quantum Physics, Optics.
%\end{description}
\end{abstract}

\keywords{Quantum Information, Quantum Physics, Optics.}
%Quantum Information
%Quantum Physics
%Optics

\maketitle

\section{Introduction}
Bosonic Gaussian states and Gaussian operations are important components in quantum information processing~\cite{Weedbrook_2012}. Despite involving an infinite-dimensional Hilbert space, they are analytically tractable and, more importantly, easy to realize in experiments. Lasers, phase-insensitive optical amplifiers, and phase-sensitive optical amplifiers all produce Gaussian states, viz., coherent states, amplified spontaneous emission (thermal) states, and squeezed states, respectively~\cite{walls2007quantum}. In addition, spontaneous parametric down conversion---the most commonly used source of optical entanglement---produces Gaussian states~\cite{walls2007quantum}. Important tasks, like quantum key distribution (QKD), can be performed with only Gaussian sources, Gaussian operations, and Gaussian measurements~\cite{grosshans2002continuous}. Gaussian attacks have also been proven to be optimum for one-way continuous-variable QKD protocols~\cite{garcia2006unconditional} and two-way continuous-variable QKD protocols~\cite{zhuang2017additive}.

However, non-Gaussian states and non-Gaussian operations are necessary for many other quantum information processing tasks, e.g., entanglement distillation~\cite{eisert2002distilling,giedke2002characterization,fiuravsek2002gaussian,zhang2010distillation}, quantum error correction~\cite{niset2009no}, optimal cloning~\cite{cerf2005non}, continuous-variable quantum computation~\cite{lloyd1999quantum,bartlett2002universal}, and cluster-state quantum computation~\cite{ohliger2010limitations,menicucci2006universal}. It has been shown that under a few reasonable assumptions, general quantum resources in the Gaussian domain cannot be distilled with Gaussian free operations~\cite{lami2018gaussian}. Moreover, non-Gaussian states and non-Gaussian operations can improve the quality of entanglement~\cite{navarrete2012enhancing} and the performance of tasks such as teleportation~\cite{opatrny2000improvement,cochrane2002teleportation,olivares2003teleportation}. For this reason,  non-Gaussian states (e.g., Fock states, N00N states~\cite{sanders1989quantum}, Schr\"{o}dinger-cat states~\cite{ourjoumtsev2006generating,ourjoumtsev2007generation}) and non-Gaussian operations (e.g., photon-number addition (PNA)~\cite{parigi2007probing,fiuravsek2009engineering,marek2008generating}, photon-number subtraction (PNS)~\cite{kitagawa2006entanglement,namekata2010non,fiuravsek2005conditional,wakui2007photon}, the qubic-phase gate~\cite{gottesman2001encoding}, the Kerr nonlinearity~\cite{nemoto2004near}, sum-frequency generation~\cite{zhuang2017optimum}, the photon-added Gaussian channels~\cite{sabapathy2017non}, and other examples~\cite{wenger2004non}) are being theoretically analyzed and experimentally realized.

An important task is thus to characterize and quantify the non-Gaussianiy (nG) utilized in each task. Quantum resource theory (QRT)~\cite{brandao2015reversible} answers this type of question. QRT has been established in various areas of physics, e.g., quantum coherence~\cite{streltsov2017colloquium,winter2016operational}, superposition~\cite{theurer17resource}, athermallity~\cite{brandao2013resource,horodecki2013fundamental}, and asymmetry~\cite{gour2008resource}. The QRT of nG is challenging because the set of Gaussian states is not convex, so the usual framework of QRT~\cite{brandao2015reversible} does not apply directly, and because of the infinite dimensional Hilbert space that is involved. Despite these difficulties, the QRT of non-Gaussian states has been developed~\cite{marian2013relative,genoni2008quantifying,genoni2010quantifying}. We explain the basic ingredients of traditional QRT via the example of non-Gaussian states: (1) resource states (non-Gaussian states), (2) free states (Gaussian states), and (3) free operations (Gaussian channels). A principal goal of QRT is to quantify the resource with a monotone---a function that maps quantum states or operations to real numbers---that satisfies three conditions: (1) zero for all free states, (2) non-zero for all resource states, and (3) non-increasing under free operations. Indeed, Refs.~\cite{marian2013relative,genoni2008quantifying} defined such a monotone based on quantum relative entropy~\cite{nielsen2002quantum,ruskai2002inequalities}, and evaluated the nG of various non-Gaussian states. 
However, the above QRT can only characterize the nG of quantum states, the nG of quantum operations is not yet well understood.

\begin{figure}
\centering
\includegraphics[width=0.4\textwidth]{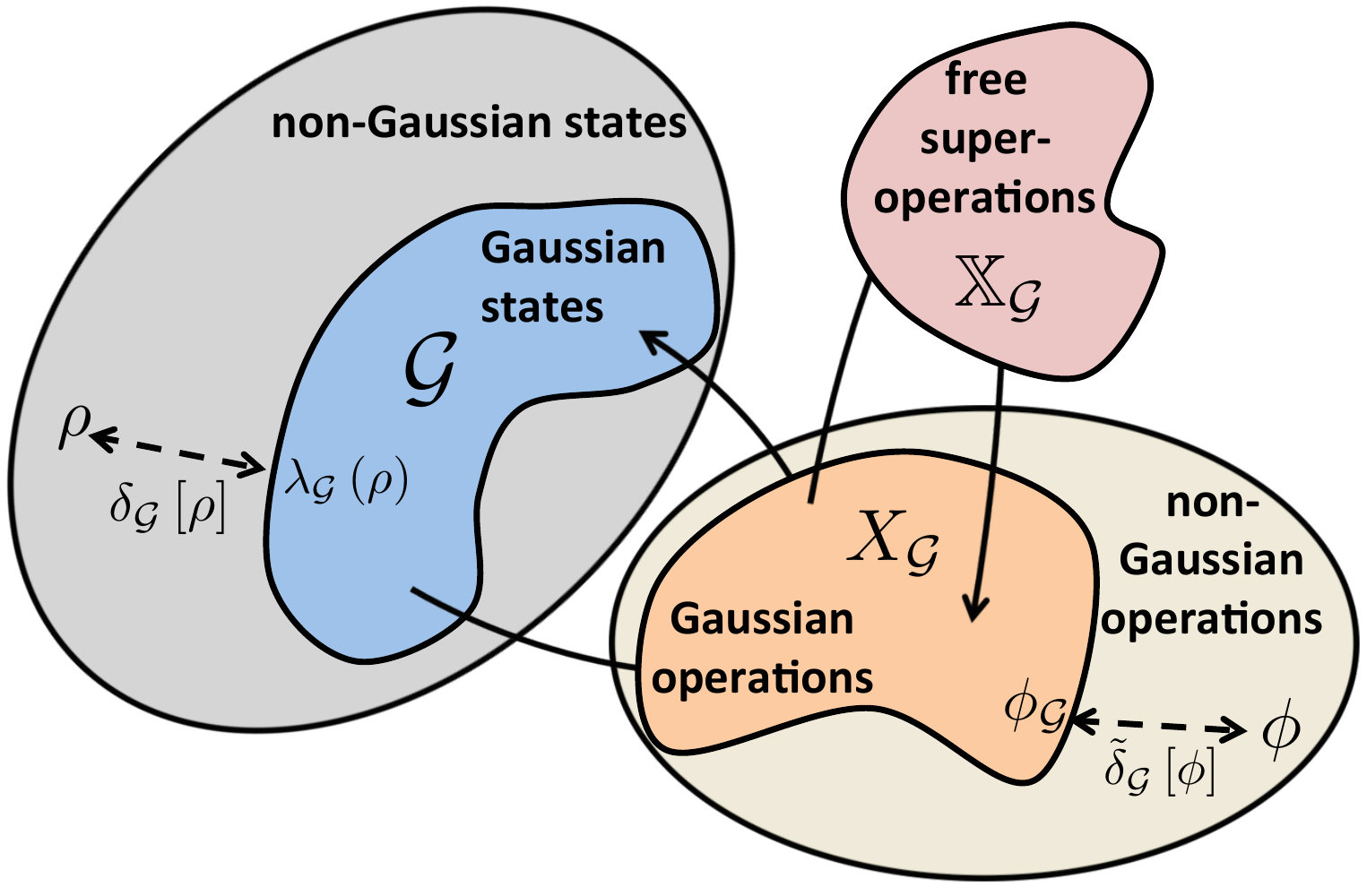}
\caption{Schematic of the resource-theory framework for non-Gaussian operations. The set of free states (Gaussian states $\calG$) is closed under the set of free operations (Gaussian operations $X_\calG$). $X_\calG$ is closed under the set of free super-operations $\mathbb{X}_\calG$. $\delta_\calG \left[\rho\right]$ is the monotone for nG of a quantum state $\rho$. It measures the difference between $\rho$ and the Gaussian state $\lambda_\calG \left(\rho\right)$ produced by the resource destroying map $\lambda_\calG$. $\tdelta_\calG \left[\phi\right]$ is the monotone for nG of a conditional quantum map $\phi$ and it also measures the deviation from some Gaussian conditional quantum map $\phi_\calG$. $\calG, X_\calG$ and $\mathbb{X}_\calG$ are non-convex.
\label{fig_frame}
}
\end{figure}

In this paper, we establish a resource theory for nG of bosonic quantum operations. In our framework, the main ingredients of QRT for quantum operations are (see the schematic in Fig.~\ref{fig_frame}): (1) resource states (non-Gaussian states) (2) free states (Gaussian states) (3) resource operations (non-Gaussian operations), (4) free operations (Gaussian operations), and (5) free super-operations (concatenation and tensoring with Gaussian channels). To quantify the nG of quantum operations, we propose a monotone---the entanglement-assisted nG generating power---that is zero for all Gaussian operations, non-zero for non-Gaussian operations, and non-increasing under free super-operations. Note that generating powers for coherence~\cite{mani2015cohering,zanardi2017measures,zanardi2017coherence,bu2017cohering,dana2017resource}, entanglement~\cite{bennett2003capacities,leifer2003optimal}, and work~\cite{navascues2015nonthermal} have been considered in other QRTs.  We also derive a lower bound and an upper bound for the monotone.  The lower bound---the generating power of nG without entanglement assistance---has been suggested in Refs.~\cite{genoni2008quantifying,genoni2010quantifying} to be a measure for nG of operations. However, it is challenging to calculate, even for unitary operations. Moreover, it is not non-increasing under the super-operation of tensoring with Gaussian channels.  

Unlike the previous suggestion, our nG monotone is analytically tractable for conditional unitary maps, including all unitary operations. As examples, we evaluate the nG of PNS and PNA. We find that the nG of both maps equals the nG of the single-photon Fock state. Our nG monotone can thus enable a quantitative characterization of nG for conditional unitary maps.
Despite the difficulty in the evaluation for general operations, we have identified two classes of operations through our nG monotone---the first class has finite nG while the second class has diverging nG. PNS and PNA are in the first class, while the binary phase-shift (BPS) channel and the Kerr nonlinearity are in the second class. For the first class, nG is finite, thus operations can be directly compared and ordered in terms of nG; for the second class, further classification may be possible by considering the rate of divergence of nG with increasing input/output mean photon number.

By utilizing the nG monotone defined in this paper and its properties, we show that all Gaussian-dilatable non-Gaussian channels defined in Ref.~\cite{sabapathy2017non,volkoff2018linear} are in the finite-nG class. The Gaussian-dilatable non-Gaussian channels are an important class of non-Gaussian channels and a starting point for our understanding of non-Gaussian operations, since their Kraus operators and input-output relations in characteristic-function form are analytically solvable. For example, this class includes the bosonic noise channel defined in Ref.~\cite{huber2017coherent}, where it has been shown that additivity violation in classical capacity is upper bounded by a constant.
It is also conjectured in Ref.~\cite{sabapathy2017non} that the set of linear bosonic channels and the set of Gaussian-dilatable channels are identical. For general bosonic channels, our result means that going beyond Gaussian-dilatable channels is important for the full understanding of non-Gaussian operations.

This paper is organized as follows. In Sec.~\ref{Sec_pre}, we introduce Gaussian states, quantum operations, and Gaussian operations, and we review the QRT of nG for non-Gaussian states. In Sec.~\ref{Sec_nG_channels}, we establish a framework for the QRT of nG for quantum operations and give the monotone, with its lower bound and upper bound. In Sec.~\ref{Sec_unitary_eva}, we evaluate the nG of two conditional unitary maps---including PNS and PNA. In Sec.~\ref{Sec_class}, we propose a classification of non-Gaussian operations. We conclude the main text in Sec.~\ref{Sec_conclusion} with discussions and future research directions. Details and proofs appear in Appendices \ref{App_QRE}-\ref{App_mixed_unitary}.

\section{Preliminaries }
\label{Sec_pre}
Here we introduce some preliminary results. In Sec.~\ref{Sec_preA}, we introduce Gaussian states; In Sec.~\ref{Sec_preB}, we introduce quantum operations; In Sec.~\ref{Sec_preC}, we introduce Gaussian operations; In Sec.~\ref{Sec_preD}, we summarize the QRT for non-Gaussian states. A complete introduction to Gaussian states and Gaussian channels can be found in Ref.~\cite{Weedbrook_2012}.

\subsection{Gaussian states}
\label{Sec_preA}
An $n$-mode bosonic continuous-variable system is described by annihilation operators $\left\{a_k, 1\le k \le n\right\}$, which satisfy the commutation relation $\left[a_k,a_j^\dagger\right]=\delta_{kj}, \left[a_k,a_j\right]=0$. One can also define real quadrature field operators $q_k=a_k+a_k^\dagger, p_k=i\left(a_k^\dagger-a_k\right)$ and formally define a real vector
${\bm x}=\left(q_1,p_1,\cdots, q_n,p_n\right)$, which satisfies the canonical commutation relation ($\hbar=2$)
$
\left[x_i,x_j\right]=2i {\bm \Omega}_{ij},
$
where ${\bm \Omega}=i \bigoplus_{k=1}^n {\bm Y} $ and ${\bm Y}$ is the Pauli matrix. A quantum state $\rho$ can be described by its Wigner characteristic function
$
\chi\left({\bm \xi}\right)=\tr \left[\rho D\left({\bm \xi}\right)\right],
$
where $\bm \xi$ is a vector of $2n$ real numbers and $
D\left({\bm \xi}\right)=\exp\left(i {\bm x}^T {\bm \Omega} {\bm \xi} \right)
$ is the Weyl operator.
A state $\rho$ is Gaussian if its characteristic function has the Gaussian form
\be
\chi \left({\bm \xi}\right)=\exp\left(-\frac{1}{2}{\bm \xi}^T \left({\bm \Omega} {\bm \Lambda } {\bm \Omega}^T\right){\bm \xi}-i \left({\bm \Omega} \bar{\bm x}\right)^T {\bm \xi}\right).
\ee
Here the $\bar{\bm x}=\braket{\bm x}_\rho$ is the state's mean and 
\be
{\bm \Lambda}_{ij}=\frac{1}{2} \braket{\{x_i-d_i,x_j-d_j\}}_\rho,
\ee
is its covariance matrix, where $\{,\}$ is the anticommutator and $\braket{A}_\rho=\tr \left(A\rho\right)$. We denote the set of normalized (i.e., unity trace) Gaussian states with $n$ modes as $\calG\left[n\right]$. The set of Gaussian states $\calG$ is the union of all $\calG\left[n\right]$, with $n\ge 1$. Any state with a non-Gaussian characteristic function is non-Gaussian.

As an example of Gaussian state, the two-mode squeezed vacuum (TMSV) state is 
\be
\ket{\zeta_\lambda}_{AA^\prime}=\sqrt{1-\lambda^2}\sum_{n=0}^\infty \lambda^n \ket{n}_A\ket{n}_{A^\prime},
\ee 
where $\ket{n}$ is a Fock state with $n$ photons. 
The covariance matrix of a TMSV can be obtained as
\begin{align}
& 
{\mathbf{{\mathbf{\Lambda}}}}_\zeta =
\left(
\begin{array}{cccc}
(2N_S+1) {\mathbf I}&2C_p{\mathbf Z}\\
2C_p {\mathbf Z}&(2N_S+1){\mathbf I}
\end{array} 
\right),
&
\end{align}
where ${\mathbf I}$, ${\mathbf Z}$ are Pauli matrices, $N_S=\lambda^2/\left(1-\lambda^2\right)$ is the mean photon number per mode, and $C_p=\sqrt{N_S\left(N_S+1\right)}$ is the phase-sensitive cross correlation.

\subsection{Quantum operations}
\label{Sec_preB}
Traditionally, a quantum operation $\calT$ is defined as a linear and completely-positive (CP) map from density operators to (unnormalized) density operators~\cite{nielsen2002quantum}. It can be expressed in terms of a unitary operator $U$ on the input in state $\rho$, and an environment $E$ in a pure state $\ket{\psi_E}$, and a projector $P$ onto $E$~\cite{nielsen2002quantum} as
\be
\calT\left(\rho\right)=\tr_E \left[\left(P\circ  U\right)\left(\rho \otimes \psi_E\right)\right].
\label{quantum_operation}
\ee
For simplicity, we have used the notation $\psi=\ket{\psi}\bra{\psi}$ to denote the density operator of a pure state $\ket{\psi}$. We also use the same notation $U$ to denote the unitary channel that applies unitary $U$ on input states, i.e.
$
U\left(\rho\right)=U \rho U^\dagger
$, and similarly $P\left(\rho\right)=P\rho P$.

When $\calT$ is also trace-preserving (TP), it is a quantum channel and can be implemented deterministically.
$\calT$ can also be non-trace-preserving. In that case, $\calT$ is implemented probabilistically. The probability of the map $\calT$ successfully happening is given by $\tr \left[\calT\left(\rho\right)\right]\le 1$ and the normalized output state is $\calT\left(\rho\right)/\tr \left[\calT\left(\rho\right)\right]$.
In various scenarios, we are interested in the enhancement provided only by the successful instances of $\calT$, e.g., when operations like PNA and PNS are used to enhance entanglement~\cite{opatrny2000improvement,cochrane2002teleportation,olivares2003teleportation,navarrete2012enhancing}. In these cases, we care more about the quantum state produced conditioned on success. Thus, we define the following post-selected completely-positive and trace-preserving (CPTP) maps.

\begin{definition}
A conditional quantum map $\phi$ takes input state $\rho$ and yields
\be
\phi\left(\rho\right)= \frac{1}{\tr \left[\calT\left(\rho\right)\right]} \calT\left(\rho\right),
\label{CMap}
\ee
where $\calT$ is a linear CP map.
\end{definition}
Map $\phi$ can be linear, when $\calT$ is TP (so $\calT$ is a quantum channel), thus conditional quantum maps include all quantum channels. Map $\phi$ can also be nonlinear, which occurs when $\calT$ is not TP, due to the normalization factor. 
The complementary map of $\phi$ is given by $\phi^c \left(\rho\right)= \calT^c \left(\rho\right)/\tr\left[\calT\left(\rho\right)\right]$, where $\calT^c$ is the complementary quantum operation and we note that $\tr \left[\calT\left(\rho\right)\right]\equiv\tr\left[ \calT^c\left(\rho\right)\right]$. In the rest of the paper, without causing confusion, we refer to conditional quantum maps as quantum operations. Note that the notion of such conditional quantum dynamics has been defined in quantum trajectory theory and quantum control~\cite{barchielli1993stochastic,wiseman1993quantum,dalvit2001unconditional,dziarmaga2004conditional,genoni2016conditional}.

In this paper we are concerned with quantum operations in infinite dimensions. We denote the set of density operators with $n$ modes as $\calH[n]$, thus we have $\calG[n]= \calH[n] \bigcap \calG$. Denote the number of input modes to channel $\phi$ as $n_\phi$ and the input Hilbert space is thus $\calH[n_\phi]$. Denote the identity operation on $\calH[n]$ as $\calI_n$. In certain cases, we will not explicitly state the dimension for simplicity ( e.g., write $\calI$ instead of $\calI_n$), as long as it does not cause any confusion.

\subsection{Gaussian operations}
\label{Sec_preC}
A quantum operation is Gaussian if it transforms Gaussian states to Gaussian states~\cite{giedke2002characterization}. Formally, the set of Gaussian operations (conditional maps) $X_\calG$ is defined as follows.

\begin{definition}
A conditional quantum map 
$ 
\phi\in X_\calG, \ \mbox{\rm iff } \forall \rho_\calG\in  \calG[n_\phi+n],\ n\in\{0,1,\cdots\}, \mbox{we have} \left(\calI_n \otimes \phi\right)\left(\rho_\calG\right)\in \calG.
$
\label{DefG}
\end{definition}
Note that if in Eq.~(\ref{CMap}) $\phi\in X_\calG$, then the original linear CP map $\calT$ is also Gaussian. And if $\phi$ is linear, the requirement in Definition~\ref{DefG} is equivalent to the weaker condition: $\forall \rho_\calG\in  \calG[n_\phi],  \mbox{we have } \phi\left(\rho_\calG\right)\in \calG$~\cite{de2015normal,de2017gaussian}.
Since on Gaussian inputs, Gaussian measurements can also be transformed to TP operations by post-processing~\cite{giedke2002characterization}, we are particularly interested in the set of Gaussian channels $X_\calG^L \subset X_\calG$. Any quantum operation outside $X_\calG$ is non-Gaussian.

All quantum channels can be extended to unitaries on the input and a vacuum environment (Stinespring dilation)~\cite{giedke2002characterization}, Gaussian unitary operations $X_\calG^U$ are therefore essential among $X_\calG$. Here we list a few Gaussian unitaries. A trivial Gaussian unitary is the identity operation $\calI_n$. Less trivial unitaries include single-mode displacement
$
D_\alpha=\exp\left(\alpha a^\dagger -\alpha^\star a \right),
$
single-mode phase rotation
$
R_\theta =\exp\left(-i \theta a^\dagger a\right),
$
single-mode squeezing
$
S_r=\exp\left[r\left(a^2-{a^\dagger}^2\right)/2\right],
$
and two-mode squeezing
$
S_{2,r}=\exp\left[-r \left(a b- a^\dagger b^\dagger\right)\right].
$
In particular, $S_{2,r}$ generates a TMSV
$
\ket{\zeta_\lambda}_{AA^\prime}
$
from vacuum inputs $\ket{0}_A\ket{0}_{A^\prime}$, i.e.,  $\left(\zeta_{\lambda}\right)_{A A^\prime}=S_{2,r}\left(0_{AA^\prime}\right)$, where $\lambda=\tanh \left(r\right)$, $\left(\zeta_{\lambda}\right)_{A A^\prime}\equiv \ket{\zeta_\lambda}_{AA^\prime} \bra{\zeta_\lambda}_{AA^\prime}$ and $0_{AA^\prime}\equiv \ket{0}_A\ket{0}_{A^\prime} \bra{0}_A\bra{0}_{A^\prime}$.

All Gaussian unitaries can be expressed as affine maps ${\bm x}\to {\bm S} {\bm x}+\Delta{\bm x}$ in the Heisenberg picture. Commutation relation preservation of $
\left[x_i,x_j\right]=2i {\bm \Omega}_{ij}
$ requires that ${\bm S}{\bm \Omega} {\bm S}^T={\bm \Omega}$, i.e., ${\bm S}$ is symplectic. In terms of the mean and covariance matrix, the affine map leads to
\be
\bar{\bm x}\to \bm  S  \bar{\bm x} +\Delta \bm x, \mbox{and } \bm  \Lambda \to \bm S \bm  \Lambda \bm S^T.
\label{Unitary_Gaussian}
\ee
Moreover, this is true regardless of whether the input state is Gaussian or not.

An arbitrary $n$-mode covariance matrix ${\bm \Lambda}$ has a symplectic diagonlization, i.e., $ \exists {\bm S}$, s.t. ${\bm S}{\bm \Omega} {\bm S}^T={\bm \Omega}$ and
$
{\bm \Lambda}={\bm S} \left(\bigoplus_{k=1}^n \mu_k {\bm I}\right) {\bm S}^T.
$ 
Here $\mu_k$'s are the eigenvalues of $i {\bm \Omega} \bm \Lambda$. Since $\mu_k \bm I$ is the covariance matrix of a thermal state with mean photon number $\left(\mu_k-1\right)/2$, this means that an arbitrary Gaussian states can be transformed into a product of thermal states with mean photon numbers $\{\left(\mu_k-1\right)/2, 1\le k \le n\}$ by a Gaussian unitary.
Thus, the entropy of such a Gaussian state
$
S\left(\rho\right)=\sum_{k=1}^n g\left(\left(\mu_k-1\right)/2\right)
$,
where $g\left(N\right)=\left(N+1\right)\log_2\left(N+1\right)-N\log_2 N$ is the entropy of a thermal state with mean photon number $N$.

As an analog to the Schmidt decomposition for finite-dimensional bipartite pure states, we have the following phase-space Schmidt decomposition~\cite{adesso2007entanglement}. Consider an arbitrary bipartite pure Gaussian state $\psi_{AB}$, with modes $\{A_k, 1\le k \le n_A\}$ and $\{B_k, 1\le k \le n_B\}$, where $n_A\le n_B$. There are local Gaussian unitaries $U_A, U_B$ that transform $\psi_{AB}$ to a tensor product of $n_A$ TMSV and $n_B-n_A$ vacuum states, i.e.,
\be
\left(U_A\otimes U_B\right)\left(\psi_{AB}\right) 
=\left[\otimes_{k=1}^{n_A} \left(\zeta_{\lambda_k}\right)_{A_kB_k}\right] \otimes \left[\otimes_{k=n_A+1}^{n_B} 0_{B_k}\right].
\label{Schmidt}
\ee

\subsection{Summary of nG resource theory for states}
\label{Sec_preD}
In a QRT, consider the set of free states to be the Gaussian states $\calG$. To characterize the nG of a quantum state $\rho$,  a relative entropy based monotone, non-increasing under free operations of Gaussian channels $X_\calG^L$, has been established~\cite{marian2013relative,genoni2008quantifying}, namely
%\ba
%\delta_\calG\left[\rho\right]&=& \min_{\rho_\calG\in \calG} S\left(\rho\| \rho_\calG\right)
%\nonumber
%\\
%&=&
%S\left(\rho\| \lambda_\calG\left(\rho\right)\right)
%\nonumber
%\\
%&=&
%S\left( \lambda_\calG\left(\rho\right)\right)-S\left(\rho\right).
%\label{nG}
%\ea
\be
\delta_\calG\left[\rho\right]= \min_{\rho_\calG\in \calG} S\left(\rho\| \rho_\calG\right)
=
S\left(\rho\| \lambda_\calG\left(\rho\right)\right)
=
S\left[ \lambda_\calG\left(\rho\right)\right]-S\left(\rho\right).
\label{nG}
\ee
Here $S\left(\rho\|\sigma\right)\equiv \tr \left[\rho \left(\log_2\rho-\log_2 \sigma\right)\right]$ is the quantum relative entropy; a brief review of its properties is given in Appendix~\ref{App_QRE}.
The first formula is a natural definition and has been shown to equal to the second formula in Ref.~\cite{marian2013relative}.
The second formula is the original proposal from Ref.~\cite{genoni2008quantifying}, and equals the third formula, where  $\lambda_\calG$ is the resource-destroying map~\cite{liu2017resource} $\rho\to \tau_\rho$, with $\tau_\rho\in \calG$ having the same mean and covariance matrix as $\rho$.
We can obtain the following lemma (proof in Appendix~\ref{proof_commute}). 

\begin{lemma}
\label{lemma_commute}
$\lambda_\calG$ commutes with any Gaussian channel $\xi_\calG\in X_\calG^L$, viz., 
$\xi_\calG\circ  \lambda_\calG =\lambda_\calG \circ \xi_\calG
$.
\end{lemma}
When Gaussian channels are considered as free operations, this condition guarantees that $S\left(\rho\| \lambda_\calG\left(\rho\right)\right)$ is a monotone~\cite{liu2017resource}. In general, however, conditional Gaussian maps do not commute with $\lambda_\calG$. A counterexample is given in Appendix~\ref{proof_commute}.

Besides continuity, $\delta_\calG\left[\cdot\right]$ satisfies the following~\cite{genoni2008quantifying}.
\begin{enumerate}[label=\text{(A\arabic*)},wide, labelwidth=!,labelindent=0pt]
\item  \label{Prop:A1} 
Non-negativity. 
$\delta_\calG \left[\rho\right]\ge 0$, with equality iff $\rho\in \calG$.

\item \label{Prop:A2} 

$\delta_\calG \left[\rho_1\otimes \rho_2\right]=\delta_\calG \left[\rho_1\right]+\delta_\calG \left[ \rho_2\right]$.

\item \label{Prop:A3}
If $\lambda_\calG\left(\rho_k\right)$'s are equal, then
$\delta_\calG \left[\sum_k p_k \rho_k\right]\le \sum_k p_k \delta_\calG \left[\rho_k\right]$.

\item \label{Prop:A4}

Invariance under a Gaussian unitary.
$\delta_\calG \left[U_\calG\rho U_\calG^\dagger\right]=\delta_\calG \left[\rho\right],\forall U_\calG \in X_\calG^U$.

\item \label{Prop:A5}

Monotonically decreasing under a partial trace.
$\delta_\calG \left[\tr_2\left(\rho_{12}\right)\right]\le \delta_\calG \left[\rho_{12}\right]$.

\item \label{Prop:A6}

Monotonically decreasing through Gaussian channels.
$\delta_\calG \left[\phi_\calG\left(\rho\right)\right]\le \delta_\calG \left[\rho\right], \forall \phi_\calG\in X_\calG^L$.

\end{enumerate}
Note that relative entropy is not superadditive in the traditional sense~\cite{zhang2013comment}.  The free set of states $\calG$ is not convex, thus precluding the results about resource state conversion in Ref.~\cite{brandao2015reversible} to hold in the resource theory of nG. Property~\ref{Prop:A6} cannot be extended to Gaussian conditional maps, a counterexample in which a Gaussian operation increases the nG of a non-Gaussian state is given in Appendix~\ref{App_counter}. This shows that even Gaussian measurements can be reduced to a Gaussian channel on Gaussian inputs by post-processing, but on non-Gaussian inputs they need to be treated differently from Gaussian channels.

\section{Resource theory of non-Gaussian operations}
\label{Sec_nG_channels}
The goal of this paper is to establish a resource theory for nG of quantum operations. We define the set of free operations to be Gaussian operations $X_\calG$. To formulate the resource theory of non-Gaussian operations, we need to find a set of super-operations that leave $X_\calG$ closed (schematic in Fig.~\ref{fig_frame}).

\begin{definition}
The set of free super-operations $\mathbb{X}_\calG$ is a set of maps that map each element in $X_\calG$ to an element in $X_\calG$. Here we consider
\be 
\mathbb{X}_\calG=\left\{\otimes \phi_\calG,\circ \phi_\calG, \phi_\calG \circ \right\},
\label{Free_operation}
\ee
which includes tensoring with a Gaussian channel ($\otimes \phi_\calG$), pre-concatenation with a Gaussian channel ($\circ \phi_\calG$) and post-concatenation with a Gaussian channel ($\phi_\calG\circ$).
\end{definition}

All the above super-operations map a Gaussian operation to another Gaussian operation. But $\mathbb{X}_\calG$ does not include general probabilistic mixing, because probabilistic mixing of Gaussian states can be non-Gaussian. We also exclude from $\mathbb{X}_\calG$ the action of taking the complement. The reason is as follows. If nG is non-increasing under taking the complement, then it must be invariant under taking the complement, because taking the complement twice gets back to the original map. However, one can construct a channel by swapping the incoming state with a non-Gaussian pure state, the channel is clearly non-Gaussian, but its complementary channel---the identity channel---is Gaussian.

The crucial step in characterizing the nG of quantum operations is to find a monotone. This monotone should be non-increasing under the set of free super-operations $\mathbb{X}_\calG$.  In Sec.~\ref{Sec_generating_power}, we will propose a monotone $\tdelta_\calG\left[\cdot\right]$ based on the entanglement-assisted generating power of quantum operations. In Sec.~\ref{lower_bound}, we obtain a lower bound $d_\calG\left[\cdot \right]$ on $\tdelta_\calG\left[\cdot\right]$. In Sec.~\ref{Sec_distance}, we obtain an upper bound $D_\calG\left[\cdot\right]$ on $\tdelta_\calG\left[\cdot\right]$ based on distance measures between quantum operations. This upper bound is in fact also a monotone. To summarize, we present two monotones, $\tdelta_\calG\left[\cdot\right]$ and $D_\calG\left[\cdot\right]$, and a lower bound $d_\calG\left[\cdot \right]$, satisfying the following relation.

\begin{theorem}
\label{theorem_relation}
For all conditional quantum maps $\phi$,
$
d_\calG\left[\phi \right]
\le 
\tdelta_\calG\left[\phi\right]
\le
D_\calG\left[\phi\right].
$
\end{theorem}
The proof is given after we introduce each quantity.
We propose $\tdelta_\calG\left[\cdot\right]$ instead of $D_\calG\left[\cdot\right]$ to be the measure of nG for quantum operations, since $\tdelta_\calG\left[\cdot\right]$ is much easier to evaluate, as we will show in Sec.~\ref{Sec_unitary_eva}. It is open whether the inequalities can be strict.

\subsection{Entanglement-assited generating power as a monotone}
\label{Sec_generating_power}
In this section, we propose a monotone for nG of quantum operations based on the entanglement-assisted generating power. 

\begin{definition}
For the input Gaussian state $\rho_{A^\prime}\in \calG[n_\phi] $ to conditional quantum map $\phi$, consider its purification $\psi_{AA^\prime}\in  \calG[2n_\phi]$. We define the entanglement-assisted nG generating power as follows
\be
\tdelta_\calG\left[\phi \right]= \max_{ \rho_{A^\prime} \in \calG[n_\phi]} \delta_\calG \left[\left(\calI_{n_\phi} \otimes \phi\right) \left(\psi_{AA^\prime}\right) \right].
\ee
\label{Def2}
\end{definition}
Before proving the properties of $\tdelta_\calG\left[\cdot\right]$ that allow it to be a monotone for nG, we justify the choice of the number of ancilla modes by the following lemma.

\begin{lemma}
\label{lemma_ancilla} $\tdelta_\calG\left[\phi \right]$ is invariant under local isometry on ancilla $A$ and giving ancilla $A$ extra modes.
\end{lemma}
The proof is based on the phase space Schmidt decomposition, details are in Appendix~\ref{App_proof_ancilla}.
In Definition~\ref{Def2}, we have chosen an ancilla with the minimum number of modes. Also, maximization over $\rho_{A^\prime}$ is equivalent to maximization over the pure state $\psi_{AA^\prime}$ due to this symmetry of purification.
This symmetry of purification also guarantees that pure states are optimum, i.e., we have an equivalent definition of $\tdelta_\calG\left[\cdot\right]$ as follows.

\begin{definition}
For $\calH\left[n+n_\phi\right]$ with $n\ge n_\phi$ modes,
\be
\tdelta_\calG\left[\phi \right]=\max_{\rho_\calG\in \calG[n+n_\phi]}\delta_\calG \left[\left(\calI_n \otimes \phi\right) \left(\rho_\calG\right) \right].
\ee
\label{Def2_equiv}
\end{definition}
This means that going to an arbitrary mixed state with an arbitrary number of modes does not increase nG. The proof that Definition~\ref{Def2_equiv} and Definition~\ref{Def2} are equivalent is as follows. By Property~\ref{Prop:A5}, we have $\delta_\calG \left[\left(\calI_n \otimes \phi \right) \left(\rho_\calG\right) \right]\le \delta_\calG \left[\left(\calI_{2n+n_\phi} \otimes \phi\right) \left(\psi_{\rho_\calG}\right) \right]$, where $\psi_{\rho_\calG}\in \calG[2n+2n_\phi]$ is the purification of $\rho_\calG\in \calG[n+n_\phi]$. Combined with symmetry of purification, we have $\max_{\rho_\calG\in \calG[n+n_\phi]}\delta_\calG \left[\left(\calI_n \otimes \phi\right) \left(\rho_\calG\right) \right]\le \max_{\psi_{\rho_\calG}\in\calG[2n+2n_\phi]}\delta_\calG \left[\left(\calI_{2n+n_\phi} \otimes \phi\right) \left(\psi_{\rho_\calG}\right) \right]= \tdelta_\calG\left[\phi \right]$. On the other hand, the reverse inequality is trivially satisfied by taking $\rho_\calG$ to be the product of the pure state in Definition~\ref{Def2} and extra vacuum ancilla.

%
%Here we make a comment about the two equivalent definition of $\tdelta_\calG\left[\cdot \right]$---\ref{Def2} and \ref{Def2_equiv}: \ref{Def2} is easier to evaluate, however, \ref{Def2_equiv} simplifies the proof of properties since it allows more general input and ancilla.

Now we give properties of $\tdelta_\calG\left[\cdot \right]$, The proofs are given in Appendix~\ref{App_proofB}.
\begin{enumerate}[label=\text{(B\arabic*)},wide, labelwidth=!,labelindent=0pt]
\item  \label{Prop:B1} 
Non-negativity. 
$
\tdelta_\calG\left[\phi \right]\ge 0, \mbox{with equality iff } \phi\in X_\calG.
$

\item  \label{Prop:B2}
Invariance under tensoring with Gaussian channels. $\forall \phi_\calG\in X_\calG^L$, we have
$
\tdelta_\calG\left[\phi \otimes \phi_\calG \right]= \tdelta_\calG\left[\phi \right].
$

\item  \label{Prop:B3}
Invariance under concatenation with a Gaussian unitary. $\forall U_\calG\in X_\calG^U$, 
$
\tdelta_\calG\left[U_\calG \circ \phi \right]=\tdelta_\calG\left[ \phi\circ U_\calG \right]= \tdelta_\calG\left[\phi \right].
$

\item \label{Prop:B4}
Monotonically decreasing under concatenation with partial trace. For $\phi$ with output $AB$, we have
$
\tdelta_\calG\left[\tr_A \circ   \phi \right]\le \tdelta_\calG\left[  \phi \right].
$

\item \label{Prop:B5}
Monotonically increasing under Stinespring dilation with a vacuum environment. Note this property is only for channels, not for general operations. Suppose $\forall \rho, \phi\left(\rho\right)=\tr_E\circ  {U_\phi\left(\rho \otimes {\bm 0}_E\right) }$, we then have
$
\tdelta_\calG\left[\phi \right]\le \tdelta_\calG\left[  U_\phi \right].
$

\item \label{Prop:B6}
Non-increasing under concatenation with a Gaussian channel. $\forall \phi_\calG\in X_\calG^L$, (1) Post-concatenation: $ \tdelta_\calG\left[\phi_\calG \circ \phi \right]\le \tdelta_\calG\left[\phi \right]$. (2) Pre-concatenation:  $\tdelta_\calG\left[\phi \circ \phi_\calG \right]\le \tdelta_\calG\left[\phi \right].
$

\item  \label{Prop:B7}
Superadditivity.	
$
\tdelta_\calG\left[\phi_1 \otimes \phi_2 \right]\ge  \tdelta_\calG\left[\phi_1\right]+\tdelta_\calG\left[\phi_2\right].
$

\end{enumerate}
It is open whether this superadditivity~\ref{Prop:B7} can be strict. Due to superadditivity, if one wants invariance under tensoring with itself, a regularization can be introduced
$
\tdelta_\calG^\infty \left[\phi \right]=\lim_{n\to \infty}  \tdelta_\calG \left[\phi^{\otimes n} \right]/n,
$ 
such that $\tdelta_\calG^\infty \left[\phi^{\otimes 2} \right]=\tdelta_\calG^\infty \left[\phi \right]$. However, unlike the case in communication capacity, where joint encoding between multiple channel uses is natural to consider; here we can simply regard $\phi$ and $\phi^{\otimes 2}$ as two different quantum operations, thus regularization is not compulsory for our resource theory.

\subsection{Generating power as a lower bound}
\label{lower_bound}
Suppose we trace out the ancilla in Definition~\ref{Def2_equiv}, we can define another function as follows.

\begin{definition}
(nG generating power)
$
d_\calG\left[\phi \right]= \max_{ \rho_\calG\in \calG[n_\phi]} \delta_{\calG}\left[\phi \left(\rho_\calG\right) \right] .
$
\label{Def1}
\end{definition}
This has been suggested in Refs.~\cite{genoni2008quantifying,genoni2010quantifying} to be a measure for the nG of quantum operations.
By considering an input in a product state with the ancilla, it is easy to see that 
$
\tdelta_\calG\left[\phi \right]\equiv d_\calG\left[\calI_{n_\phi} \otimes \phi \right]\ge d_\calG\left[\phi \right],
$
by Property~\ref{Prop:A5}. Thus the first part of Theorem~\ref{theorem_relation} is true.
If the above inequality can be strict (which seems plausible), because the identity $\calI_{n_\phi}$ is a Gaussian channel, we cannot prove invariance nor non-increasing under tensoring with Gaussian channels. Moreover, $d_\calG\left[\phi \right]= 0$ only implies $ \forall \rho_\calG\in \calG, \phi \left(\rho_\calG\right) \in \calG$, which does not necessarily mean $\phi\in X_\calG$ according to Definition~\ref{DefG}. Thus, it only satisfies Properties~\ref{Prop:B3}-\ref{Prop:B7} (see Appendix~\ref{App_proofC} for details). Additionally, it is difficult to calculate $d_\calG\left[\cdot \right]$ even for unitary operations, since it requires maximization over mixed states and the entropy of a non-Gaussian mixed state is difficult to calculate. In contrast, $\tdelta_\calG$ can be analytically evaluated, as we will show in Sec.~\ref{Sec_unitary_eva}.

\subsection{Upper bound---distance as a monotone}
\label{Sec_distance}
Another natural definition for the nG of quantum operations can be obtained from a geometric approach. Since the diamond norm~\cite{kitaev2002classical} is difficult to calculate, here we introduce the following.

\begin{definition}
Consider conditional quantum maps $\phi_1$ and $\phi_2$ each with the $n$ input modes. We define a
measure for their difference by 
\be 
D_\calG\left(\phi_1,\phi_2\right)
=\max_{\psi_\calG\in\calG[2n]}
S\left[\left(\calI_{n} \otimes \phi_1\right) \left(\psi_\calG\right)\|\left(\calI_{n} \otimes \phi_2\right) \left(\psi_\calG\right)\right],
\ee 
which is equivalent to
\be 
D_\calG\left(\phi_1,\phi_2\right)=\max_{\rho_\calG\in\calG[n]}
S\left[\left(\calI_{n} \otimes \phi_1\right) \left(\rho_\calG\right)\|\left(\calI_{n} \otimes \phi_2\right) \left(\rho_\calG\right)\right].
\ee 
\label{D_def}
\end{definition}
In the first formula, we have restricted the state to be pure and within $\calG[2n]$. An argument similar to Lemma~\ref{lemma_ancilla}'s proof gives the second formula.
Now, one can define a measure of nG by the the distance from the closest Gaussian conditional map with the same number of input modes.

\begin{definition}
(nG distance)
$
D_\calG\left[\phi\right]\equiv \min_{\phi_\calG\in X_\calG}D_\calG\left(\phi,\phi_\calG\right).
$
\label{D}
\end{definition}
Now we show that the second part of Theorem~\ref{theorem_relation} is true. We will not explicitly state the dimension in the following proof for simplicity.
\ba
&&D_\calG\left[\phi\right]=\min_{\phi_\calG\in X_\calG}\max_{\psi_\calG\in\calG}S\left[\left(\calI \otimes \phi\right) \left(\psi_\calG\right)\|
\left(\calI \otimes \phi_\calG\right) \left(\psi_\calG\right)\right]
\nonumber
\\
&&\ge 
\max_{\psi_\calG\in\calG}
\min_{\phi_\calG\in X_\calG}
S\left[\left(\calI \otimes \phi\right) \left(\psi_\calG\right)\|
\left(\calI \otimes \phi_\calG\right) \left(\psi_\calG\right)\right]
\nonumber
\\
&&\ge \max_{\psi_\calG\in\calG}
\min_{\rho_\calG\in \calG}
S\left[\left(\calI \otimes \phi\right) \left(\psi_\calG\right)\|\rho_\calG\right]
\nonumber
\\
&&=\max_{\psi_\calG\in\calG}\delta_\calG \left[ \left(\calI \otimes \phi\right) \left(\psi_\calG\right)\right]=\tdelta_\calG\left[\phi \right].
\nonumber
\ea
The first inequality is due to the max-min inequality~\cite{boyd2004convex}, the second inequality is due to the fact that $\left(\calI \otimes \phi_\calG\right) \left(\psi_\calG\right)\in \calG$, and the last equality is due to Eq.~(\ref{nG}) and Definition~\ref{Def2}.

We can show that $D_\calG\left[\cdot\right]$ satisfies Properties~\ref{Prop:B1}-\ref{Prop:B6}, which qualifies it to be a measure of nG for quantum operations (see Appendix~\ref{App_proofD} for details). Moreover, we can show that it satisfies	
$
D_\calG\left[\phi_1 \otimes \phi_2 \right]\ge \max\left( D_\calG\left[\phi_1\right], D_\calG\left[\phi_2\right]\right).
$
It is open whether this can be improved to superadditivity.

\section{Example: Conditional unitary maps}
\label{Sec_unitary_eva}

We now introduce conditional unitary maps.

\begin{definition}
A conditional quantum map is a conditional unitary map if it is one-to-one and maps all pure states to pure states.
\end{definition}
Conditional unitary maps include  unitary operations, like the single-mode self-Kerr unitary~\cite{genoni2010quantifying}, and operations like PNA and PNS~\cite{kim2008scheme,navarrete2012enhancing}. For a conditional unitary map $U$, because the output-ancilla is jointly pure when the input-ancilla is pure, combining Eq.~(\ref{nG}) and Definition~\ref{Def2} gives
\ba
\tdelta_\calG\left[U \right]
&=&
\max_{ \rho_{A^\prime}\in \calG} S\left[ \lambda_\calG\left[\left(\calI \otimes U\right) \left(\psi_{AA^\prime}\right)\right]\right].
\ea
For fixed $\rho_{A^\prime}$, $S\left[ \lambda_\calG\left(\left(\calI \otimes U\right) \left(\psi_{AA^\prime}\right)\right)\right]$ can be analytically obtained by calculating the entropy of the Gaussian state $\lambda_\calG\left[\left(\calI \otimes U\right) \left(\psi_{AA^\prime}\right)\right]$, which can be obtained from its covariance matrix. Moreover, the Gaussian state $\rho_{A^\prime}$ being maximized over can be fully characterized by its mean and covariance matrix. Thus, the overall maximization can be solved analytically without too much difficulty.
For example, in the single-mode case, the general input-ancilla state
\be
\ket{\psi_{\alpha,\theta,r,\lambda}}_{AA^\prime} = D_\alpha R_\theta  S_r \ket{\zeta_\lambda}_{AA^\prime}
\label{general_input}
\ee
only depends on four parameters---the  displacement $\alpha$, phase rotation $\theta$, squeezing $r$, and two-mode squeezing $\lambda$. Note here that $D_\alpha, R_\theta$ and $S_r$ act on the input $A^\prime$. 

Below, we consider two specific single-mode conditional maps---the PNS $\phi_{\rm PNS}$ and PNA $\phi_{\rm PNA}$---and evaluate their nG's analytically. For simplicity, we consider the ideal $\phi_{\rm PNS}$ and $\phi_{\rm PNA}$, which are described by the annihilation and creation operators $a$ and $a ^\dagger$~\cite{kim2008scheme,navarrete2012enhancing}. Experimental schemes of PNS and PNA can be found in Refs.~\cite{parigi2007probing,fiuravsek2009engineering,marek2008generating,kitagawa2006entanglement,namekata2010non,fiuravsek2005conditional,wakui2007photon}. Both $\phi_{\rm PNS}$ and $\phi_{\rm PNA}$ are one-to-one and produce a pure state when the input is pure, thus they are conditional unitary maps.

{\em Photon-number subtraction.---}
When the input and ancilla are in the joint state given by Eq.~(\ref{general_input}), the joint state of the output and ancilla  is
$
\ket{\psi}_{AB}=N_{\rm PNS} a_B\ket{\psi_{\alpha,\theta,r,\lambda}}_{AB}
$,
where the normalization factor is
$
N_{\rm PNS} =\left(|\alpha|^2+\left(\left(1+2N_S\right)\cosh \left(2r\right)-1\right)/2\right)^{-1/2}.
$
Because of Property~\ref{Prop:A4}, $\ket{\psi}_{AB}$ has the same nG as
\begin{align}
&\ket{\xi}_{AB}=S_r^\dagger R_\theta ^\dagger  D_\alpha^\dagger \ket{\psi}_{AB}
\nonumber
\\
&= N_{\rm PNS}  \left(e^{-i \theta}\left(\cosh\left(r\right) a_B-\sinh\left(r\right)a_B^\dagger \right)+\alpha\right)\ket{\zeta_\lambda}_{AB},
\label{xi_AB}
\end{align}
where $\ket{\xi}_{AB}$ is a superposition of photon-number added TMSV, photon-number subtracted TMSV, and TMSV, so it is non-Gaussian. By changing the global phase properly, we can choose $\alpha>0$. 

To calculate the covariance matrix of $\xi_{AB}$, we consider the expectation values of operators $X\in \left\{a_A, a_B, a_A^2, a_B^2, a_A^\dagger a_A, a_B^\dagger a_B, a_A^\dagger a_B,a_A a_B\right\}$, which can be found from
\begin{align}
&\braket{X}_{\xi_{AB}}\equiv 
\bra{\xi}_{AB} X\ket{\xi}_{AB}
=N^2\left\{
\alpha^2 \braket{X}_{\zeta_\lambda}
\right.
\nonumber
\\
&
+\alpha e^{-i\theta}\cosh\left(r\right)\braket{Xa_B}_{\zeta_\lambda}
-\alpha e^{-i\theta}\sinh\left(r\right)\braket{Xa_B^\dagger}_{\zeta_\lambda}
\nonumber
\\
&
+\alpha e^{i\theta} \cosh\left(r\right)\braket{a_B^\dagger X}_{\zeta_\lambda}
-\alpha e^{i\theta} \sinh\left(r\right)\braket{a_B X}_{\zeta_\lambda}
\nonumber
\\
&
+\cosh^2\left(r\right)\braket{a_B^\dagger X a_B}_{\zeta_\lambda}
+\sinh\left(r\right)^2 \braket{a_B X a_B^\dagger}_{\zeta_\lambda}
\nonumber
\\
&
\left.
-\frac{1}{2}\sinh\left(2r\right) \left(
\braket{a_B X a_B}_{\zeta_\lambda}
+\braket{a_B^\dagger X a_B^\dagger}_{\zeta_\lambda}
\right)
\right\}.
\label{X_sub}
\end{align}
Since TMSV $\zeta_\lambda$ has zero mean, each term can be solved by Gaussian moment factoring. The covariance matrix can be obtained by the method in Appendix~\ref{App_Cov}, however the expression is too lengthy to display here. With the covariance matrix in hand, the entropy can be obtained easily by the method in Sec.~\ref{Sec_preC}.

After the maximization over $r,\alpha,\theta,N_S$, we find that 
\be
\tdelta_\calG\left[\phi_{\rm PNS} \right]=\delta_\calG\left[\ket{1}\bra{1}\right]=2,
\ee
which is achieved by $\alpha=0$ and arbitrary $r,\theta,N_S$.
This result equals the lower bound $d^\prime_\calG$ obtained in Ref.~\cite{genoni2007measure} for the special case of $N_S=0,\alpha=0$.

{\em Photon-number addition.---}
The nG analysis for PNA parallels what we have done for PNS. The joint state of the output and ancilla is
$
\ket{\psi}_{AB}=N_{\rm PNA}  a_B^\dagger\ket{\psi_{\alpha,\theta,r,\lambda}}_{AB}
$,
where
$
N_{\rm PNA}=\left(|\alpha|^2+\left(\left(1+2N_S\right)\cosh \left(2r\right)+1\right)/2\right)^{-1/2}.
$
Because of Property~\ref{Prop:A4}, $\ket{\psi}_{AB}$ has the same nG as
\begin{align}
&\ket{\xi}_{AB}=S_r^\dagger R_\theta^\dagger  D_\alpha^\dagger \ket{\psi}_{AB}
\nonumber
\\
&= N_{\rm PNA} \left(e^{i \theta}\left(\cosh\left(r\right) a_B^\dagger-\sinh\left(r\right)a_B \right)+\alpha^\star\right)\ket{\zeta_\lambda}_{AB}.
\label{xi_AB_2}
\end{align}
Intuitively, since it is again a superposition of photon-number added TMSV, photon-number subtracted TMSV and TMSV, the maximum nG should be the same as that of $\phi_{PNS}$. However, because $\cosh\left(r\right)\ge \sinh\left(r\right)$, the parameter space here is slightly different. This difference can be dealt with by realizing that the new  expectation values can be obtained by exchanging $-\sinh \left(r\right)$ with $\cosh\left(r\right)$ and $\theta$ with $-\theta$ in Eq.~(\ref{X_sub}) (fixing $\alpha>0$), and using the new normalization factor.

After the maximization over $r,\alpha,\theta,N_S$, we find that 
\be
\tdelta_\calG\left[\phi_{\rm PNA} \right]=\tdelta_\calG\left[\phi_{\rm PNS} \right]=\delta_\calG\left[\ket{1}\bra{1}\right]=2,
\ee
which is achieved by $\alpha=0$ and arbitrary $r,\theta,N_S$.

\section{Classification---finite nG and diverging nG}
\label{Sec_class}
\begin{figure}
\centering
\includegraphics[width=0.45\textwidth]{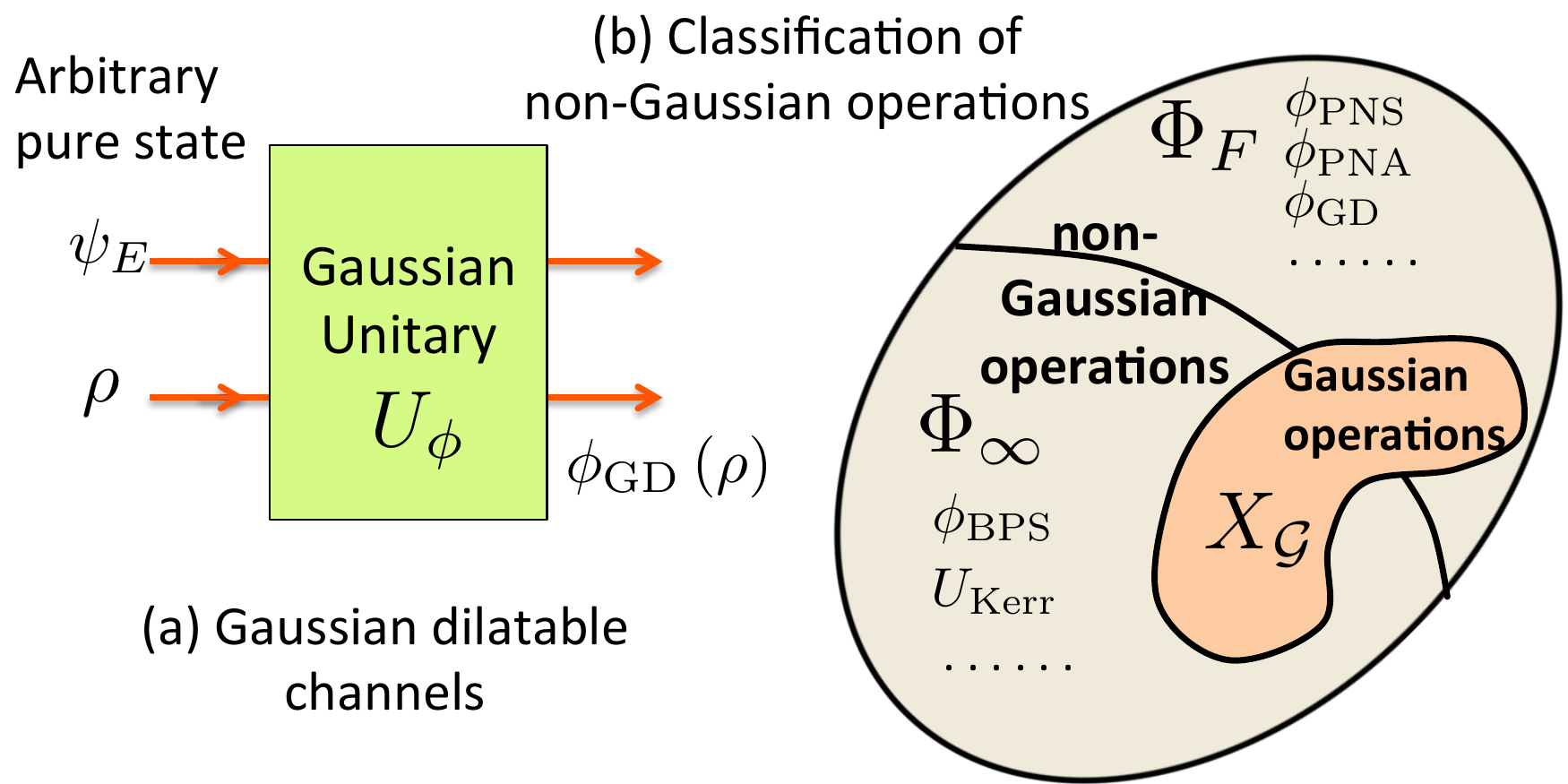}
\caption{
(a) Schematic of a Gaussian-dilatable channel $\phi_{\rm GD}$, $\psi_E$ is the environment in an arbitrary pure state. (b) Schematic of the classification of non-Gaussian operations into: (1) finite-nG class $\Phi_F$, including $\phi_{\rm PNS}$, $\phi_{\rm PNA}$ and Gaussian-dilatable non-Gaussian channels $\phi_{\rm GD}$, and (2) diverging-nG class $\Phi_\infty$, including the binary phase-shift channel $\phi_{\rm BPS}$ and the self-Kerr unitary $U_{\rm Kerr}$. 
\label{fig_2}
}
\end{figure}

In the above examples, nG is finite. However, for other quantum operations there is a potential divergence caused by the infinite dimensionality of states in $\calG$. Consider Definition~\ref{Def2}. If the overall output energy is bounded by $N_S$, then
$ 
\tdelta_\calG\left[\phi \right]\le \max_{ \rho_{A^\prime}\in \calG[n_\phi]} S\left( \lambda_\calG\left(\rho_{AB}\right)\right)\le n g\left(N_S/n\right).
$
The factor $n$ is the total number of modes in the output and ancilla.
Since $g\left(N_S\right)\sim \log_2 N_S$, when $N_S\gg1$, the growth rate of $\tdelta_\calG\left[\phi \right]$ with the allowed output energy is at most logarithmic.
It may be tempting to constrain the input/output energy in Definition~\ref{Def2} to define an energy-constrained version of generating power. However, because concatenation of Gaussian channels can change the energy constraint on the input/output of the original conditional quantum map, such constraints will invalidate Properties~\ref{Prop:B3} and~\ref{Prop:B6}. So an energy-constrained generating power is not a meaningful monotone for nG.

Based on the above observation, we classify non-Gaussian operations into two classes (schematic in Fig.~\ref{fig_2}(b)). The first class of operations has finite $\tdelta_\calG$ despite allowing the input to have infinite energy. We denote this class of operations $\Phi_F$. 

\begin{definition} Finite-nG class.
\be
\Phi_F=\left\{{\rm conditional \ quantum \ map \ }\phi \mid  0 <\tdelta_\calG\left[\phi\right]<\infty \right\}.
\ee
\end{definition} 
As we have shown in Sec.~\ref{Sec_unitary_eva}, PNA and PNS both belong to this class, i.e.,
\be
\phi_{\rm PNS}\in \Phi_F,\phi_{\rm PNA}\in \Phi_F.
\ee
For operations in $\Phi_F$, we can compare and rank their nG based on the $\tdelta_\calG\left[\phi \right]$ value.

The second class of operations has diverging $\tdelta_\calG$, when the output energy increases. We denote this class of operations as $\Phi_\infty$. 

\begin{definition} Diverging-nG class.
\be
\Phi_\infty=\left\{{\rm conditional \ quantum \ map \ }\phi \mid  \tdelta_\calG\left[\phi\right]=\infty \right\}.
\ee
\end{definition} 

To identify the diverging-nG class, it is often useful to consider the lower bound
\be
\tdelta_\calG\left[\phi \right]\ge d_\calG \left[\phi\right]\ge \delta_\calG \left[\phi\left(\ket{\alpha}\bra{\alpha}\right)\right],
\label{ineq_lb}
\ee
where the coherent state $\ket{\alpha}$ is the input to the map. 
If we can show that $\delta_\calG \left[\phi\left(\ket{\alpha}\bra{\alpha}\right)\right]$ diverges to $\infty$ as $|\alpha|^2$ increases, then we can conclude that $\phi\in \Phi_\infty$.

In the following, we give more examples of operations in $\Phi_F$ and $\Phi_\infty$.

{\em Gaussian-dilatable channels.---} In Ref.~\cite{sabapathy2017non}, a class of non-Gaussian channels called Gaussian-dilatable non-Gaussian channels is introduced. A channel is Gaussian-dilatable if it has a Stinespring dilation composed of a Gaussian unitary $U_\phi\in X_\calG^U$ and an ancilla in a fixed pure state $\psi_E$ with finite energy (schematic in Fig.~\ref{fig_2}(a)). A Gaussian-dilatable channel $\phi_{\rm GD}$'s output on arbitrary input $\rho$ can be written as
\be
\phi_{\rm GD}\left(\rho\right)=\tr_E \left(U_\phi \left(\rho \otimes \psi_E\right)\right).
\ee
All Gaussian channels are trivially Gaussian-dilatable. $\phi_{\rm GD}$ is non-Gaussian when $\psi_E$ is non-Gaussian.
For Gaussian-dilatable channels, the output's characteristic function can be analytically obtained from the input's characteristic function and the Kraus operators are also analytically attainable. Thus, Gaussian-dilatable channels are an important starting point for the study of non-Gaussian channels and operations. For example, it includes the bosonic noise channel defined in Ref.~\cite{huber2017coherent}, where it has been shown that its additivity violation in classical capacity is upper bounded by a constant. 
It is also conjectured in Ref.~\cite{sabapathy2017non} (see Conjecture 1 in the reference) that the set of linear bosonic channels and the set of Gaussian-dilatable channels are identical.

The nG of a Gaussian-dilatable channel satisfies
\ba
\tdelta_\calG\left[\phi_{\rm GD} \right]&=&\max_{\psi_\calG} \delta_\calG \left[\left(\calI_{n_\phi} \otimes \left(\tr_E \circ U_\phi \right)\right)\left(\psi_\calG \otimes \psi_E\right)\right]
\nonumber
\\
&\le& 
\max_{\psi_\calG} \delta_\calG \left[\left(\calI_{n_\phi} \otimes U_\phi\right) \left(\psi_\calG \otimes \psi_E\right)\right]
\nonumber
\\
&=&\max_{\psi_\calG} \delta_\calG \left(\psi_\calG \otimes \psi_E\right)=\delta_\calG \left[\psi_E\right],
\ea
where the first inequality is from Property~\ref{Prop:A5}, the second equality is from Property~\ref{Prop:A4} and the last equality is from Property ~\ref{Prop:A2}. Because the nG of the state $\psi_E$ is finite and does not depend on the input or output, we immediately have the following theorem.

\begin{theorem}
\label{theorem_GD}
Every Gaussian-dilatable non-Gaussian channel is in the finite-nG class, i.e.,
\be
\phi_{\rm GD}\in \Phi_F.
\ee
\end{theorem}
The fact that $\tdelta_\calG\left[\phi_{\rm GD} \right]\le \delta_\calG \left[\psi_E\right]$ is intuitive, since all nG of this channel comes from the non-Gaussian environment and all other operations are Gaussian. Here we have considered an ancilla with finite energy. An ancilla with infinite energy is only meaningful when one considers a sequence of ancilla with increasing finite energy. However, the ancilla of a fixed channel cannot depend on the energy of the input state, thus in terms of the growth with the input energy, the amount of nG is bounded for Gaussian dilatable channels~\footnote{In principle, one can encode all possible output states into an ancilla with infinite energy, thus considering an infinite-energy ancilla is not meaningful.}.
Note that our argument does not rule out the possibility that all channels might be approximately Gaussian-dilatable. The formulation of approximate Gaussian-dilatable channels still requires more work.

{\em Binary phase-shift channel.---}Consider a single-mode channel that applies a phase shift $R_\pi$ with probability $1/2$, i.e.
\be
\phi_{\rm BPS}\left(\rho\right)=\frac{1}{2}\rho+\frac{1}{2} R_\pi \rho R_\pi^\dagger.
\ee
Let the input be a coherent state $\ket{\alpha}$ ($\alpha>0$), so that the mean and covariance matrix of the output $\phi_{\rm BPS}\left(\ket{\alpha}\bra{\alpha}\right)=\frac{1}{2}\ket{\alpha}\bra{\alpha}+\frac{1}{2} \ket{-\alpha}\bra{-\alpha}$ are $\left(0,0\right)$ and ${\rm Diag}\left(4\alpha^2+1,1\right)$. The entropy of the Gaussian state with the same mean and covariance matrix is $g\left(\left(\sqrt{4\alpha^2+1}-1\right)/2\right)$, while $S\left(\phi_{\rm BPS}\left(\ket{\alpha}\bra{\alpha}\right)\right)\le1$. Thus we have 
$
\delta_\calG \left[\phi_{\rm BPS}\left(\ket{\alpha}\bra{\alpha}\right)\right]\ge
g\left(\left(\sqrt{4\alpha^2+1}-1\right)/2\right)-1,
$ and equality is achieved as $\alpha\to \infty$. It is diverging as $\alpha$ increases. Thus $
\delta_\calG \left[\phi_{\rm BPS}\left(\ket{\alpha}\bra{\alpha}\right)\right]
$ diverges as $\alpha$ increases, so
\be
\phi_{\rm BPS}\in \Phi_\infty.
\ee
In fact, if one considers the input and ancilla to be in a TMSV, it is straightforward to show (details in Appendix~\ref{App_mixed_unitary}) that
$
\tdelta_\calG \left[\phi_{\rm BPS}\right]\ge 2 g\left(N_S/2\right)-1,
$
when the output and ancilla have total energy constraint $N_S$. Thus the rate of divergence is $\log_2\left(N_S\right)$, which is the maximum rate of divergence.

{\em Self-Kerr unitary.---}
Consider now the single-mode self-Kerr unitary
\be
U_{\rm Kerr}=\exp\left(-i \gamma \left(a^\dagger a\right)^2\right).
\ee
The lower bound $
\delta_\calG \left[U_{\rm Kerr}\left(\ket{\alpha}\bra{\alpha}\right)\right]$ has been found to diverge maximally, as $\log_2\left(N_S\right)$, where $N_S=|\alpha|^2$~\cite{genoni2010quantifying}. So we have
\be
U_{\rm Kerr}\in \Phi_\infty.
\ee

We have classified non-Gaussian operations into two classes $\Phi_F$ and $\Phi_\infty$. Within the class $\Phi_F$, nG is finite and thus comparing and ordering different operations is straightforward. Within the class $\Phi_\infty$, even though all nG are infinite, they can have different rates of divergence. So classification based on those rates is possible.

It is an open question whether all linear maps (quantum channels) in $\Phi_F$ are Gaussian-dilatable. If it is true, then because of Theorem~\ref{theorem_GD}, it  would imply that the class of Gaussian-dilatable non-Gaussian channels and the class of finite-nG channels are equal. It is also open whether there is a minimum set of operations in $\Phi_F$, such that any other operations in $\Phi_F$ can be simulated by this set of operations and Gaussian operations in $X_\calG$, in terms of the generation of non-Gaussian states from Gaussian inputs.

\section{Conclusions}
\label{Sec_conclusion}

Gaussian states and Gaussian operations are inadequate for various tasks, such as universal quantum computing, entanglement distillation, and quantum error correction. So non-Gaussian states and operations are naturally considered as resources for these tasks. A quantum resource theory for nG in states and operations is a starting point for understanding the utility of nG.

In this paper, we extended the resource theory of non-Gaussian states in Refs.~\cite{marian2013relative,genoni2008quantifying,genoni2010quantifying} to non-Gaussian operations and established a monotone to quantify the amount of nG. This monotone can be analytically calculated for conditional unitary maps like PNS and PNA. We also provided a lower bound and an upper bound for this monotone to assist in the calculation and analysis of nG.

More importantly, our monotone enables us to classify non-Gaussian operations into (1) the finite-nG class, and (2) the diverging-nG class. Within the first class, nG is finite, thus direct comparison and ordering of operations is straightforward. Within the second class, nG diverges as the output energy increases. Further classification may be possible through comparing rates of divergence.

We gave several examples of quantum operations in each class. In particular, we showed that all Gaussian-dilatable non-Gaussian channels are in the finite-nG class. Thus, not all non-Gaussian channels are Gaussian-dilatable. Gaussian-dilatable channels are important because their properties, such as their Kraus operators, are relatively easy to obtain, making them a starting point for studying of non-Gaussian channels and operations. For example, recent results~\cite{huber2017coherent} show that the non-additivity violation in a bosonic noise channel, which is Gaussian-dilatable, is mild. However, our results suggest that focusing on Gaussian-dilatable channels is not enough for the full understanding of non-Gaussian channels.

An important future research direction is the operational resource theory of non-Gaussian operations, like the one for coherence~\cite{winter2016operational}. For example, how to quantify the power of different non-Gaussian operations for specific tasks, like quantum computation and entanglement distillation, is worthy of investigation. This problem is also related to channel simulation in terms of production of non-Gaussian states. One can also ask whether there is a finite set of universal non-Gaussian operations, such that all non-Gaussian states can be produced by this set of non-Gaussian operations and arbitrary Gaussian operations starting from Gaussian states. The answer is yes, because universal quantum computation is possible with Gaussian operations plus one single non-Gaussian operation~\cite{lloyd1999quantum}. However, it is not clear whether the class of finite-nG operations can enable universal quantum computing or it is necessary to have operations from the diverging-nG class.

Another important future task is the further classification of non-Gaussian operations. As an analog, there are bound entanglement states~\cite{horodecki1998mixed} that have zero distillable entanglement, and cannot be directly used to enhance teleportation. Similarly, a mixture of Gaussian channels, e.g., the BPS channel, seems less useful than the Kerr nonlinearity for many tasks such as universal computation, while they are both in the diverging-nG class with the same rate of divergence. A more delicate classification, based on the convex resource theory of non-Gaussianity~\cite{convex1,convex2}, that distinguishes these two types of non-Gaussian operations is an important step towards the full classification of non-Gaussian operations. 

\begin{acknowledgments}
QZ thanks Zi-Wen Liu and Ryuji Takagi for discussions. QZ and JHS are supported by the Air Force Office of
Scientific Research Grant No. FA9550-14-1-0052. QZ also acknowledges the Claude E. Shannon Research
Assistantship.  PWS is supported by the National Science
Foundation under Grant No. CCF-1525130 and National Science
Foundation through the Science and Technology Centers for Science of Information under Grant No. CCF0-939370.
\end{acknowledgments}

\appendix

\section{Properties of quantum relative entropy}
\label{App_QRE}
The relative entropy of two quantum states $\rho$ and $\sigma$ is defined as
$
S\left(\rho\|\sigma\right)\equiv \tr \left[\rho \left(\log_2\rho-\log_2 \sigma\right)\right].
$
Besides continuity, it has the following properties~\cite{nielsen2002quantum,ruskai2002inequalities}.
\begin{enumerate}[label=\text{(O\arabic*)},wide, labelwidth=!,labelindent=0pt]
\item  
Non-negativity (Klein's inequality). $S\left(\rho\|\sigma\right)\ge0$.
\item  
Joint convexity.
\vspace{-3ex}
\ba S\left(p\rho_1+\left(1-p\right)\rho_2\|p\sigma_1+\left(1-p\right)\sigma_2\right)
\nonumber
\\
\le p S\left(\rho_1\|\sigma_1\right)+\left(1-p\right)S\left(\rho_2\|\sigma_2\right).
\nonumber
\ea

\vspace{-3ex}
\item 
Monotonically decreasing under partial trace.
\vspace{-2ex}
\be
S\left(\tr_2\rho_{12}\|\tr_2\sigma_{12}\right)\le S\left(\rho_{12}\|\sigma_{12}\right).
\nonumber
\ee

\vspace{-3ex}
\item 
Monotonically decreasing under quantum operation. 
$
S\left(\varepsilon\left(\rho\right)\|\varepsilon\left(\sigma\right)\right)\le S\left(\rho\|\sigma\right).
$
Equal when $\varepsilon$ is an isometry.

\item 
Additivity of product states.

$S\left(\rho_1\otimes \rho_2\|\sigma_1\otimes \sigma_2\right)=S\left(\rho_1\|\sigma_1\right)+S\left(\rho_2\| \sigma_2\right)$.

\item 
$
2S\left(\rho_{12}\|\sigma_{12}\right)\ge S\left(\rho_1\|\sigma_1\right)+S\left(\rho_2\|\sigma_2\right).
$
Superadditivity can be established by a better multiplicative constant~\cite{capel2017superadditivity}.

\end{enumerate}
\section{Proof of Lemma~\ref{lemma_commute}}
\label{proof_commute}

\begin{proof} 
A Gaussian channel $\xi_\calG$ can be extended to a Gaussian unitary on its input and an environment~\cite{giedke2002characterization,Weedbrook_2012}, which can be expressed as a linear transform on the mean and covariance matrix in Eq.~(\ref{Unitary_Gaussian}). The output can be obtained by tracing out part of the joint output of this Gaussian unitary. Thus $\xi_\calG$ produces a state (not necessarily Gaussian) with mean and covariance matrix $\left(\bar{\bm x}^\prime, \bm \Lambda^\prime\right)$ as function of the mean and covariance matrix $\left(\bar{\bm x},\bm \Lambda\right)$ of the input $\rho$.
So $\left(\lambda_\calG \circ \xi_\calG\right) \left(\rho\right)$ is a Gaussian state with mean and covariance matrix $\left(\bar{\bm x}^\prime, \bm \Lambda^\prime\right)$. On the other hand, $\left(\xi_\calG\circ  \lambda_\calG\right) \left(\rho\right)$ is also a Gaussian state with mean and covariance matrix $\left(\bar{\bm x}^\prime, \bm \Lambda^\prime\right)$. Since a Gaussian state is uniquely specified by its mean and covariance matrix, we have $\left(\xi_\calG\circ  \lambda_\calG\right) \left(\rho\right)=\left(\lambda_\calG \circ \xi_\calG\right) \left(\rho\right), \forall \rho$.
\end{proof}

A counterexample for the generalization to conditional Gaussian maps is constructed here. Consider the conditional map 
\be
\calT_\alpha\left(\rho_{AA^\prime}\right)=\frac{\braket{\alpha |_{A^\prime}\rho_{AA^\prime}|\alpha}_{A^\prime}}{\tr_A\braket{\alpha |_{A^\prime}\rho_{AA^\prime}|\alpha}_{A^\prime}}, 
\label{T_counter}
\ee
which projects on $A^\prime$ and outputs $A$, where $\ket{\alpha}_{A^\prime}$ is the coherent state with amplitude $\alpha>0$. Consider the input $\sigma_{AA^\prime}=\left(\ket{\alpha}_A\bra{\alpha}\otimes \ket{\alpha}_{A^\prime}\bra{\alpha}+\ket{-\alpha}_A\bra{-\alpha}\otimes \ket{-\alpha}_{A^\prime}\bra{-\alpha}\right)/2$. In the following, we show that $\left(\lambda_\calG \circ \calT_\alpha\right)\left(\sigma_{AA^\prime}\right) $ and $\left(\calT_\alpha \circ \lambda_\calG\right)\left(\sigma_{AA^\prime}\right)$ have different means and are thus different Gaussian states. We have that
\be
\calT_\alpha\left(\sigma_{AA^\prime}\right)=\frac{1}{1+e^{-4\alpha^2}}\left(\ket{\alpha}_A \bra{\alpha}+e^{-4\alpha^2}\ket{-\alpha}_A \bra{-\alpha}\right),
\ee
where expectation value is
\be 
\braket{a}_{\calT_\alpha\left(\sigma_{AA^\prime}\right)}=\frac{1-e^{-4\alpha^2}}{1+e^{-4\alpha^2}}\alpha.
\label{exp_val_1}
\ee

From results in Appendix~\ref{App_Cov}, the mean of $\sigma_{AA^\prime}$ is $\left(0,0,0,0\right)$, and its covariance matrix is
\begin{align}
& 
{\mathbf{{\mathbf{\Lambda}}}}_\sigma =
\left(
\begin{array}{cccc}
4\alpha^2+1&0&4\alpha^2&0\\
0&1&0&0\\
4\alpha^2 &0&4\alpha^2+1&0\\
0&0&0&1
\end{array} 
\right).
&
\end{align}
The density matrix of $\lambda_\calG \left(\sigma_{AA^\prime}\right)$ can be obtained through the $P$-function~\cite{Weedbrook_2012} as
\be
\lambda_\calG \left(\sigma_{AA^\prime}\right)=\int_{-\infty}^\infty d\alpha_1 \frac{1}{\sqrt{2\pi}\alpha}e^{-\frac{\alpha_1^2}{2\alpha^2}}\ket{\alpha_1}_A\bra{\alpha_1}\otimes \ket{\alpha_1}_{A^\prime}\bra{\alpha_1}.
\ee
The output of the map is 
$
\left(\calT_\alpha \circ \lambda_\calG\right) \left(\sigma_{AA^\prime}\right)\propto \int_{-\infty}^\infty d\alpha_1
$
$
\exp{\left(-\frac{\alpha_1^2}{2\alpha^2}-\left(\alpha_1^2+\alpha^2-2\alpha_1 \alpha\right)\right)}
\ket{\alpha_1}_A\bra{\alpha_1}. 
$ It is then straightforward to see that
\be
\braket{a}_{\calT_\alpha \circ \lambda_\calG \left(\sigma_{AA^\prime}\right)}=\frac{2\alpha^3}{1+\alpha^2},
\ee
which is not equal to $\braket{a}_{\lambda_\calG\circ \calT_\alpha\left(\sigma_{AA^\prime}\right)}=\braket{a}_{\calT_\alpha\left(\sigma_{AA^\prime}\right)}$ given in Eq.~(\ref{exp_val_1}) for finite $\alpha>0$.

\section{Counterexample}
\label{App_counter}

Consider a non-Gaussian state
$
\rho_{AA^\prime}\simeq \sqrt{\epsilon} \ket{\alpha}_{A^\prime}\bra{\alpha}\otimes \ket{n}_A\bra{n}+\sqrt{1-\epsilon}\ket{-\alpha}_{A^\prime}\bra{-\alpha}\otimes \rho_A,
$
where $\epsilon\ll1$, $\alpha\gg1$ and $\rho_A\in \calG$. We have $\delta_\calG \left[\rho_{AA^\prime}\right]\ll 1$ from continuity. For the Gaussian conditional map in Eq.~(\ref{T_counter}), we have
$
\calT_\alpha\left(\rho_{AA^\prime}\right)\simeq \ket{n}_A\bra{n}.
$
This means that $\delta_\calG\left[\calT_\alpha\left(\rho_{AA^\prime}\right)\right]\simeq \delta_\calG\left[\ket{n}_A\bra{n}\right]\gg \delta_\calG \left[\rho_{AA^\prime}\right]$, i.e., nG can increase under a Gaussian conditional map.

\section{Proof of Lemma~\ref{lemma_ancilla}}
\label{App_proof_ancilla}
\begin{proof}
We use methods similar to those in Ref.~\cite{WatrousJ2011}. Any pure Gaussian state $\psi_{AA^\prime}$, with $A$ having $n\ge n_\phi$ modes and $A^\prime$ having $n_\phi$ modes, has phase-space Schmidt decomposition (Eq.~(\ref{Schmidt}) in main text)
\be
U_{A}\left(\psi_{AA^\prime}\right) 
=\left[\otimes_{k=n_\phi+1}^{n} 0_{A_k}\right]\otimes \left[U_{A^\prime}^\dagger \left(\otimes_{k=1}^{n_\phi} \left(\zeta_{\lambda_k}\right)_{A_kA_k^\prime}\right)\right].
\nonumber
\ee
Thus, $U_A$ can allow a Gaussian isometry $u_A$ from $\calH[n]$ to $\calH[n_\phi]$ such that $\psi_{AA^\prime}=\left( \left(u_{A}^{-1} \circ u_{A}\right) \otimes \calI_{n_\phi} \right) \left( \psi_{AA^\prime}\right)$.

Now let $\psi_{A_\phi A^{\prime  }}=\left( u_{A} \otimes \calI_{n_\phi}\right) \left(\psi_{AA^\prime}\right)\in \calG[2n_\phi]$. Due to relative entropy's invariance under isometries, $u_A\in X_\calG^L$, and Lemma~\ref{lemma_commute}, we get
\begin{align} 
&\delta_\calG \left[\left(\calI_n \otimes \phi \right) \left(\psi_{AA^\prime}\right) \right]  
\nonumber
\\
&
=
S\left[\left(\calI_n \otimes \phi\right) \left(\psi_{AA^\prime}\right) \| \lambda_\calG\left(\left(\calI_n \otimes   \phi\right) \left(\psi_{AA^\prime}\right) \right)\right]
\nonumber
\\
&
=
S\left[\left(u_A \otimes \phi\right) \left(\psi_{AA^\prime}\right) \| u_A\circ \lambda_\calG\left(\left(\calI_n \otimes   \phi\right) \left(\psi_{AA^\prime}\right) \right)\right]
\nonumber
\\
&
=
S\left[\left(\calI_{n_\phi} \otimes \phi\right) \left(\psi_{A_\phi A^\prime}\right) \|  \lambda_\calG\left(\left(u_A \otimes   \phi\right) \left(\psi_{AA^\prime}\right) \right) \right]
\nonumber
\\
&
=\delta_\calG \left[\left(\calI_{n_\phi} \otimes \phi \right) \left(\psi_{A_\phi A^\prime}\right) \right]. 
\label{proof_dim}
\end{align}
\end{proof}

\section{Proofs of properties \ref{Prop:B1}-\ref{Prop:B7}}
\label{App_proofB}
In most proofs we use Definition~\ref{Def2_equiv} as a starting point, and we will simplify the notation for the domain of maximization, e.g., writing ${\rho_\calG\in \calG[n+n_\phi]}$ as $\rho_\calG\in\calG$. Also, we will not explicitly state the dimension of the identity operator $\calI$ when it's not necessary.
\begin{enumerate}[label=\text{(B\arabic*)},wide, labelwidth=!,labelindent=0pt]
\item   
\begin{proof}
Non-negativity follows directly from Property~\ref{Prop:A1}. If $\phi\in X_\calG$, it is easy to see that $\tdelta_\calG\left[\phi \right]=0$ since $\calI \otimes \phi\in X_\calG$.
Now we prove the reverse part. Suppose $\tdelta_\calG\left[\phi \right]=0$, then by Definition~\ref{Def2_equiv}, $\forall \rho_\calG\in \calG$, we have $\delta_{\calG}\left[\left(\calI \otimes \phi\right)\left(\rho_\calG\right) \right]=0$. By Definition~\ref{DefG} and Property~\ref{Prop:A1}, we get $\phi\in X_\calG$.
\end{proof}

\item  
\begin{proof}
(1) $\tdelta_\calG\left[\phi \otimes \phi_\calG \right]
=
\max_{\rho_\calG\in \calG}
$
$
\delta_\calG \left[\left(\calI \otimes \phi \otimes \phi_\calG \right)\left(\rho_\calG\right) \right]
$
$
\ge 
\max_{\rho_\calG^\prime\in \calG}
$
$
\delta_\calG \left[\left(\calI \otimes \phi \right) \left(\rho_\calG^\prime \right) \right]
=\tdelta_\calG\left[\phi \right].
$
The inequality is obtained by taking trace over the output of $\phi_\calG$ and using Property~\ref{Prop:A5}.

(2) 
$\tdelta_\calG\left[\phi \otimes \phi_\calG \right]
=
\max_{\rho_\calG\in \calG}\delta_\calG \left[\left(\calI \otimes \phi \otimes \phi_\calG\right) \left(\rho_\calG\right) \right]
$
$
=
\max_{\rho_\calG\in \calG}\delta_\calG \left[\left(\calI \otimes \phi \otimes \calI_{\phi_\calG}\right) \left(\calI\otimes \calI_{n_\phi} \otimes \phi_\calG\right)\left(\rho_\calG\right)  \right]
$
$
\le 
\max_{\rho_\calG^\prime\in \calG}
$
$\delta_\calG \left[\left(\calI \otimes \phi \otimes \calI_{n_{\phi_\calG}}\right) \left(\rho_\calG^\prime \right) \right]=\tdelta_\calG\left[\phi \right],
$
where the inequality follows since $\calI\otimes \calI_{n_\phi} \otimes \phi_\calG\left(\rho_\calG\right)\in \calG$ and in the last equality we have used the symmetry of purification in Lemma~\ref{lemma_ancilla}.
\end{proof}

\item  
\begin{proof} (1) From Property~\ref{Prop:A4}, 
$ \tdelta_\calG\left[U_\calG \circ \phi \right]$
$=\max_{\rho_\calG\in \calG}
$
$
\delta_\calG \left[\left(\calI \otimes U_\calG\right) \circ \left(\calI \otimes \phi\right) \left(\rho_\calG\right) \right]
$
$
= \max_{\rho_\calG\in\calG}
$
$
\delta_\calG \left[ \left(\calI \otimes \phi\right) \left(\rho_\calG\right) \right]=\tdelta_\calG\left[\phi \right].
$

\noindent(2) $\tdelta_\calG\left[\phi \circ U_\calG \right]=\max_{\rho_\calG\in\calG}\delta_\calG \left[ \left(\calI \otimes \phi\right) \circ \left(\calI \otimes U_\calG\right) 
\right.
$
$
\left.
\left(\rho_\calG\right) \right]
$
$
= \max_{\rho_\calG\in\calG}\delta_\calG \left[ \left(\calI \otimes \phi\right) \left(\rho_\calG\right) \right]=\tdelta_\calG\left[\phi \right]$, where we have used $\left(\calI \otimes U_\calG\right)\left(\calG\right)=\calG$.
\end{proof}

\item 
\begin{proof}
$\tdelta_\calG\left[\tr_A \circ\phi \right]
=\max_{\rho_\calG\in \calG}
$
$
\delta_\calG \left[\left(\calI \otimes \left(\tr_A \circ \phi\right)\right) \right.
$
$
\left.
\left(\rho_\calG\right) \right]
=\max_{\rho_\calG\in \calG}
$
$
\delta_\calG \left[\tr_A \left(\left(\calI \otimes  \phi \right)\left(\rho_\calG\right) \right)\right]
$
$
\le 
\max_{\rho_\calG\in \calG}
$
$  
\delta_\calG \left[\left(\calI \otimes  \phi\right) \left(\rho_\calG\right) \right]
=\tdelta_\calG\left[  \phi \right]$. The inequality follows from Property~\ref{Prop:A5}.
\end{proof}

\item 
\begin{proof}
$
\tdelta_\calG\left[\phi \right]
$
$
=\max_{\rho_\calG\in \calG}
$
$
\delta_\calG \left[\left(\calI \otimes \left(\tr_E\circ  U_\phi\right)\right)
\right.
$
$
\left.
\left(\rho_\calG \otimes {\bm 0}_E\right)  \right]
$
$
\le
\max_{\rho_\calG^\prime \in \calG}
$
$
\delta_\calG \left[\left(\calI \otimes   \left(\tr_E \circ U_\phi\right)\right)\left(\rho_\calG^\prime\right)  \right]
=
\tdelta_\calG\left[\tr_E \circ U_\phi \right]
\le 
\tdelta_\calG\left[  U_\phi \right].
$
The first inequality is due to expanding the set of states over which the maximization is performed. The second inequality is because of Property~\ref{Prop:B4}.
\end{proof}
			
\item 
\begin{proof} 
%From property~\ref{Prop:B2},~\ref{Prop:B3} and~\ref{Prop:B4}, $\tdelta_\calG\left[ \phi \right]=\tdelta_\calG\left[ \calI_{n_\phi}\otimes \phi \right]=\tdelta_\calG\left[ U_\calG \circ \left(\calI_{n_\phi}\otimes \phi \right) \right]\ge \tdelta_\calG\left[ \tr_A \left( U_\calG \circ \left(\calI_{n_\phi}\otimes \phi \right) \right)\right]=\tdelta_\calG\left[\phi_\calG \circ \phi \right]$. We used the fact that $\phi_\calG$ can be extended to a unitary $U_\calG$ over input $A^\prime$ and ancilla $A$. Similarly, $\tdelta_\calG\left[ \phi \right]=\tdelta_\calG\left[ \calI_{n_\phi}\otimes \phi \right]=\tdelta_\calG\left[ \left(\calI_{n_\phi}\otimes \phi \right)\circ U_\calG \right]\ge \tdelta_\calG\left[ \tr_A \left(  \left(\calI_{n_\phi}\otimes \phi \right)\circ U_\calG \right)\right]=\tdelta_\calG\left[\phi \circ \phi_\calG \right]$ 
(1) From Property~\ref{Prop:A6},
$ \tdelta_\calG\left[\phi_\calG \circ \phi \right]$
$=\max_{\rho_\calG\in \calG}
$
$
\delta_\calG \left[\left(\calI \otimes \phi_\calG\right) \circ \left(\calI \otimes \phi\right) \left(\rho_\calG\right) \right]
$
$
\le \max_{\rho_\calG\in\calG}
$
$
\delta_\calG \left[ \left(\calI \otimes \phi\right) \left(\rho_\calG\right) \right]=\tdelta_\calG\left[\phi \right].
$

\noindent (2) $\tdelta_\calG\left[\phi \circ \phi_\calG \right]=\max_{\rho_\calG\in\calG}\delta_\calG \left[ \left(\calI \otimes \phi\right) \circ \left(\calI \otimes \phi_\calG\right) 
\right.
$
$
\left.
\left(\rho_\calG\right) \right]
$
$
\le \max_{\rho_\calG\in\calG}\delta_\calG \left[ \left(\calI \otimes \phi\right) \left(\rho_\calG\right) \right]=\tdelta_\calG\left[\phi \right]$. The inequality uses the fact that $\left(\calI \otimes \phi_\calG\right) \left(\rho_\calG\right)\in \calG$.
\end{proof}

\item  
\begin{proof}
In Definition~\ref{Def2_equiv}, choose the ancilla to be in $\calH[n_{\phi_1}] \otimes \calH[n_{\phi_2}]$, so we can write $\calI=\calI_1 \otimes \calI_2$, where $\calI_k$ is the identity operator on $\calH[n_{\phi_k}]$, thus
$
\tdelta_\calG\left[\phi_1 \otimes \phi_2 \right]=\max_{\rho_\calG\in \calG[2n_{\phi_1}+2n_{\phi_2}]
}\delta_\calG \left[\left(\calI_1 \otimes \phi_1 \otimes \calI_2 \otimes \phi_2\right) \left(\rho_\calG\right) \right]
\ge 
\max_{\rho_1 \in  \calG[2n_{\phi_1}]}\max_{\rho_2 \in  \calG[2n_{\phi_2}]}\delta_\calG \left[\left(\calI_1 \otimes \phi_1 \otimes \calI_2 \otimes \phi_2 \right)
\right.
$
$
\left.
\left(\rho_1 \otimes \rho_2\right) \right]
= \tdelta_\calG\left[\phi_1\right]+\tdelta_\calG\left[\phi_2\right]
$, where in the last step we used Property~\ref{Prop:A2}.
\end{proof}
\end{enumerate}

\section{Properties of $d_
\calG$}
\label{App_proofC} 
\begin{enumerate}[label=\text{(C\arabic*)},wide, labelwidth=!,labelindent=0pt]

\item  \label{Prop:C2}
Invariance under concatenation with a Gaussian unitary. $\forall U_\calG\in X_\calG^U$, we have
$
d_\calG\left[U_\calG \circ \phi \right]=d_\calG\left[ \phi\circ U_\calG \right]= d_\calG\left[\phi \right].
$
  
\begin{proof}
(1) $d_\calG\left[U_\calG \circ \phi \right]= \max_{ \rho_\calG\in \calG} \delta_{\calG}\left[U_\calG \left(\phi \left(\rho_\calG \right)\right) \right]= \max_{ \rho_\calG\in \calG} \delta_{\calG}\left[\phi \left(\rho_\calG \right) \right]=d_\calG\left[\phi \right]$, where we used Property~\ref{Prop:A4}.
(2) $d_\calG\left[\phi \circ U_\calG \right]=\max_{ \rho_\calG\in \calG} \delta_{\calG}\left[\phi \circ U_\calG \left(\rho_\calG\right) \right]= \max_{ \rho_\calG\in \calG} \delta_{\calG}\left[\phi \left(\rho_\calG\right) \right]= d_\calG\left[\phi \right]$, where we have used $U_\calG \left(\calG\right)= \calG$.
\end{proof}

\item  \label{Prop:C3}
Monotonically decreasing under the concatenation with partial trace. For $\phi$ with output $AB$,
we have
$
d_\calG\left[\tr_A \circ   \phi \right]\le d_\calG\left[  \phi \right].
$

\begin{proof}
$d_\calG\left[\tr_A \circ\phi \right]
=\max_{\rho_\calG\in \calG}\delta_\calG \left[ \tr_A \circ \phi \left(\rho_\calG\right) \right]
=\max_{\rho_\calG\in \calG}\delta_\calG \left[\tr_A \left( \phi \left(\rho_\calG\right) \right)\right]\le \max_{\rho_\calG\in \calG}
$
$  
\delta_\calG \left[ \phi \left(\rho_\calG\right) \right]
=d_\calG\left[  \phi \right]$. The inequality follows from Property~\ref{Prop:A5}.
\end{proof}

\item  \label{Prop:C4}
Monotonically increasing under Stinespring dilation with a vacuum environment. Note this property is only for channels, not for general operations. Suppose $\forall \rho, \phi\left(\rho\right)=\tr_E\circ  {U_\phi\left(\rho \otimes {\bm 0}_E\right) }$, we have
$
d_\calG\left[\phi \right]\le d_\calG\left[  U_\phi \right].
$

\begin{proof}
$
d_\calG\left[\phi \right]
=\max_{\rho_\calG\in \calG}
\delta_\calG \left[ \tr_E\circ  U_\phi\left(\rho_\calG \otimes {\bm 0}_E\right)  \right]
$
$
\le
\max_{\rho_\calG^\prime\in \calG}
$
$
\delta_\calG \left[ \tr_E\circ  U_\phi\left(\rho_\calG^\prime\right)  \right]
=d_\calG\left[\tr_E\circ  U_\phi \right]\le d_\calG\left[U_\phi \right]
$
The first inequality is due to expanding the set of states over which the maximization is performed. The second inequality is from Property~\ref{Prop:C3}.
\end{proof}

\item   \label{Prop:C5}
Non-increasing under concatenation with a Gaussian channel. $\forall \phi_\calG\in X_\calG^L$,
(1) Post-concatenation: 
$
d_\calG\left[\phi_\calG \circ \phi \right]\le d_\calG\left[\phi \right].
$
(2) Pre-concatenation: 
$ 
d_\calG\left[\phi \circ \phi_\calG \right]\le d_\calG\left[\phi \right].
$

\begin{proof}
(1)	$d_\calG\left[\phi_\calG \circ \phi \right]= \max_{ \rho_\calG\in \calG} \delta_{\calG}\left[\phi_\calG \left(\phi \left(\rho_\calG \right)\right) \right]\le \max_{ \rho_\calG\in \calG} \delta_{\calG}\left[\phi \left(\rho_\calG \right) \right]=d_\calG\left[\phi \right]$, where we used Property~\ref{Prop:A6}.
(2) $d_\calG\left[\phi \circ \phi_\calG \right]=\max_{ \rho_\calG\in \calG} \delta_{\calG}\left[\phi \circ \phi_\calG \left(\rho_\calG\right) \right]\le \max_{ \rho_\calG\in \phi_\calG \left(\calG\right)} \delta_{\calG}\left[\phi \left(\rho_\calG\right) \right]\le d_\calG\left[\phi \right]$, where we have used $\phi_\calG \left(\calG\right)\subset \calG$.
\end{proof}

\item  \label{Prop:C6}
Superadditivity.	
$
d_\calG\left[\phi_1 \otimes \phi_2 \right]\ge  d_\calG\left[\phi_1\right]+d_\calG\left[\phi_2\right].
$

\begin{proof}
$
d_\calG\left[\phi_1 \otimes \phi_2 \right]
=\max_{\rho_\calG\in  \calG[n_{\phi_1}+n_{\phi_2}]}
$
$
\delta_\calG \left[ \left(\phi_1 \otimes  \phi_2\right) \left(\rho_\calG\right) \right]
\ge 
\max_{\rho_1 \in  \calG[n_{\phi_1}]}
$
$
\max_{\rho_2 \in  \calG[n_{\phi_2}]}
$
$
\delta_\calG \left[ \left(\phi_1 \otimes  \phi_2 \right)
\right.
$
$
\left.
\left(\rho_1 \otimes \rho_2\right) \right]
= d_\calG\left[\phi_1\right]+d_\calG\left[\phi_2\right]
$, where in the last step we used Property~\ref{Prop:A2}.
\end{proof}

\end{enumerate}

\section{Properties of $D_\calG$}
\label{App_proofD}
\begin{enumerate}[label=\text{(D\arabic*)},wide, labelwidth=!,labelindent=0pt]
\item  \label{Prop:D1} 
Non-negativity. 
$
D_\calG\left[\phi\right]\ge 0, \mbox{with equality iff } \phi\in X_\calG.
$
		
\begin{proof}
This follows from Property~\ref{Prop:B1} of $\tdelta_\calG\left[\phi \right]$ and Theorem~\ref{theorem_relation}. Alternatively, this result can be obtained from the non-negativity of quantum relative entropy.
\end{proof}

\item  \label{Prop:D2} 
Invariance under tensoring with a Gaussian channel. $\forall \xi_\calG\in X_\calG^L$, we have 
$
D_\calG\left[\phi \otimes \xi_\calG \right]= D_\calG\left[\phi \right].
$

\begin{proof}
(1) First we prove $D_\calG\left[\phi \otimes \xi_\calG\right]\ge D_\calG\left[\phi \right]$. 
\begin{align}
&D_\calG\left[\phi \otimes \xi_\calG \right]
\nonumber
\\
&
=
\min_{\phi_\calG\in X_\calG}\max_{\rho_\calG\in\calG}
S\left[
\left(\calI \otimes \phi\otimes \xi_\calG\right) \left(\rho_\calG\right)\|
\left(\calI \otimes \phi_\calG\right) \left(\rho_\calG\right)\right]
\nonumber
\\
&
\ge 
\min_{\phi_\calG\in X_\calG}\max_{\rho_\calG^\prime\in\calG}S\left[
\calI \otimes \phi\otimes \xi_\calG \left(\rho_\calG^\prime \otimes \rho_1\right)\|
\calI \otimes \phi_\calG \left(\rho_\calG^\prime \otimes \rho_1\right)\right]
\nonumber
\\
&
= 
\min_{\phi_\calG\in X_\calG}\max_{\rho_\calG^\prime\in\calG}
S\left[
\calI \otimes \phi \left(\rho_\calG^\prime \right)\otimes \xi_\calG\left(\rho_1\right)\|
\calI \otimes \phi_\calG \left(\rho_\calG^\prime \otimes \rho_1\right)\right]
\nonumber
\\
&
\ge 
\min_{\phi_\calG\in X_\calG}\max_{\rho_\calG^\prime\in\calG}S\left[
\left(\calI \otimes \phi\right) \left(\rho_\calG^\prime \right)\|
\calI \otimes \left( \tr_{\xi_\calG}\circ \phi_\calG\right) \left(\rho_\calG^\prime \otimes \rho_1\right)\right]
\nonumber
\\
&
= 
\min_{\phi_\calG^\prime\in X_\calG}\max_{\rho_\calG^\prime\in\calG}S\left[
\left(\calI \otimes \phi\right) \left(\rho_\calG^\prime\right)\|
\left(\calI \otimes \phi_\calG^\prime \right) \left(\rho_\calG^\prime\right)\right]=D_\calG\left[\phi \right].
\end{align}
The first inequality is from limiting the maximization to states of the form $\rho_\calG^\prime \otimes \rho_1$.
The second inequality is from relative entropy's monotonically decreasing under partial trace. 
The last equality is because $\forall \phi_\calG \in X_\calG$, $\left( \tr_{\xi_\calG}\circ \phi_\calG\right)$ is a Gaussian operation that takes input $\sigma$ and outputs to $\calH[n_\phi]$, and every Gaussian operation with the same input/output dimension with $\phi$ can be extended to a another Gaussian operation by trivially tensoring with the identity. 

(2) Now we prove $D_\calG\left[\phi \otimes \xi_\calG\right]\le D_\calG\left[\phi \right]$. 
\begin{align}
&D_\calG\left[\phi \otimes \xi_\calG \right]
\nonumber
\\
&
=\min_{\phi_\calG\in X_\calG}\max_{\rho_\calG\in\calG}S\left[\left(\calI \otimes \phi\otimes \xi_\calG\right) \left(\rho_\calG\right)\|
\left(\calI \otimes \phi_\calG\right) \left(\rho_\calG\right)\right]
\nonumber
\\
&\le 
\min_{\phi_\calG^\prime\in X_\calG}\max_{\rho_\calG\in\calG}
S\left[\left(\calI \otimes \phi\otimes \xi_\calG\right) \left(\rho_\calG\right)\|
\left(\calI \otimes \phi_\calG^\prime \otimes \xi_\calG\right) \left(\rho_\calG\right)\right]
\nonumber
\\
&\le 
\min_{\phi_\calG^\prime\in X_\calG}\max_{\rho_\calG\in\calG}S\left[
\left(\calI^\prime \otimes \phi\right) \left(\rho_\calG\right)\|
\left(\calI^\prime \otimes \phi_\calG^\prime \right) \left(\rho_\calG\right)\right]=D_\calG\left[\phi \right].
\end{align}
The first inequality is due to limiting the minimization to operations of the form $\phi_\calG^\prime \otimes \xi_\calG$.
The last inequality is due to relative entropy's monotonically decreasing under quantum operations and symmetry in the ancilla.
\end{proof}

\item  \label{Prop:D3} 
Invariance under concatenation with a Gaussian unitary. $\forall U_\calG\in X_\calG^U$, we have
$
D_\calG\left[U_\calG \circ \phi \right]=D_\calG\left[ \phi\circ U_\calG \right]= D_\calG\left[\phi \right].
$  

\begin{proof}
(1) $U_\calG$ has inverse $U_\calG^{-1}$. 
So 
\begin{align}
&D_\calG\left[U_\calG \circ \phi\right]
\nonumber
\\
&=\min_{\phi_\calG\in X_\calG}\max_{\psi_\calG\in\calG}
S\left[
\left(\calI \otimes \left(U_\calG \circ\phi \right)\right) \left(\psi_\calG\right)\|
\left(\calI \otimes \phi_\calG\right) \left(\psi_\calG\right)\right]
\nonumber
\\
&=
\min_{\phi_\calG\in X_\calG}\max_{\psi_\calG\in\calG}
S\left[
\left(\calI \otimes \phi\right) \left(\psi_\calG\right)\|
\left(\calI \otimes \left(U_\calG^{-1} \circ\phi_\calG \right)\right) \left(\psi_\calG\right)\right]
\nonumber
\\
&=\min_{\phi_\calG^\prime\in X_\calG}\max_{\psi_\calG\in\calG}S\left[
\left(\calI \otimes \phi\right) \left(\psi_\calG\right)\|
\left(\calI \otimes \phi_\calG^\prime\right)  \left(\psi_\calG \right)\right]=D_\calG\left[ \phi\right].
\end{align}
We have used the invariance of relative entropy under isometries.

(2) $\forall \phi_\calG\in X_\calG$, let $\phi_\calG^\prime=\phi_\calG \circ U_\calG^{-1}\in X_\calG$.  
\begin{align}
&D_\calG\left[ \phi\circ U_\calG\right]
\nonumber
\\
&=\min_{\phi_\calG\in X_\calG}\max_{\psi_\calG\in\calG}S\left[
\left(\calI \otimes \left( \phi \circ U_\calG \right)\right) \left(\psi_\calG\right)\|
\left(\calI \otimes \phi_\calG\right) \left(\psi_\calG\right)\right]
\nonumber
\\
&=
\min_{\phi_\calG^\prime\in X_\calG}
\max_{\psi_\calG\in\calG}
S\left[
\calI \otimes \left( \phi \circ U_\calG \right) \left(\psi_\calG\right)\|
\calI \otimes \left( \phi_\calG^\prime \circ U_\calG \right) 
\left(\psi_\calG\right)\right]
\nonumber
\\
&=\min_{\phi_\calG^\prime\in X_\calG}\max_{\psi_\calG^\prime\in\calG}S\left[
\left(\calI \otimes  \phi\right)  \left(\psi_\calG^\prime\right)\|
\left(\calI \otimes  \phi_\calG^\prime\right)  \left(\psi_\calG^\prime\right)\right]
=D_\calG\left[ \phi\right].
\end{align}
We have used $U_\calG\left(\calG\right)=\calG$.
\end{proof}

\item  \label{Prop:D4} 
Monotonically decreasing under concatenation with a partial trace. For $\phi$ with output $AB$, we have
$
D_\calG\left[\tr_A \circ   \phi \right]\le D_\calG\left[  \phi \right].
$

\begin{proof}
$D_\calG\left[\tr_A \circ   \phi \right]$
\vspace{-0.5em}
\ba
&&=\min_{\phi_\calG\in X_\calG}\max_{\psi_\calG\in\calG}
S\left[
\left(\calI \otimes \left(\tr_A \circ\phi \right)\right) \left(\psi_\calG\right)
\|
\left(\calI \otimes \phi_\calG\right) \left(\psi_\calG\right)\right]
\nonumber
\\
&&\le \min_{\phi_\calG^\prime \in X_\calG}\max_{\psi_\calG\in\calG}
S\left[
\calI \otimes \left(\tr_A \circ\phi \right) \left(\psi_\calG\right)\|\calI \otimes \left(\tr_A \circ\phi_\calG^\prime \right) \left(\psi_\calG\right)\right]
\nonumber
\\
&&\le\min_{\phi_\calG^\prime\in X_\calG}\max_{\psi_\calG\in\calG}S\left[
\left(\calI \otimes \phi\right) \left(\psi_\calG\right)
\|
\left(\calI \otimes \phi_\calG^\prime \right) \left(\psi_\calG \right)\right]
\nonumber
\\
&&
=D_\calG\left[ \phi\right].
\ea
The first inequality is due to limiting to minimization over $\phi_\calG$ that can be written as $\tr_A \circ \phi_\calG^\prime$. The second inequality is due to relative entropy's monotonically decreasing under a partial trace.
\end{proof}

\item   \label{Prop:D5} 
Monotonically increasing under Stinespring dilation with a vacuum environment. Note this property is only for channels, not for general operations.  Suppose $\forall \rho, \phi\left(\rho\right)=\tr_E\circ  {U_\phi\left(\rho \otimes {\bm 0}_E\right) }$, then
$
D_\calG\left[\phi \right]\le D_\calG\left[  U_\phi \right].
$

\begin{proof}
We have
\begin{align}
&
D_\calG\left[\phi \right]
\nonumber
\\
&
=\min_{\phi_\calG\in X_\calG}
\max_{\psi_\calG\in\calG}
S\left[
\left(
\calI \otimes \left(\tr_E \circ U_\phi \right) 
\right)
\left(\psi_\calG \otimes {\bm 0}_E \right)\|
\left(
\calI \otimes \phi_\calG
\right) \left(\psi_\calG\right)\right]
\nonumber
\\
&
\le 
\min_{\phi_\calG^\prime \in X_\calG}
\max_{\psi_\calG\in\calG}
S\left[
\left(
\calI \otimes \left(\tr_E \circ U_\phi \right)
\right)
\left(\psi_\calG\otimes {\bm 0}_E\right)
\right.
\nonumber
\\
& \ \ \ \ \ \ \ \ \ \ \ \ \ \ \ \ \ \ \ \ \ \ \ \ \ \ \
\left.
\|
\left(
\calI \otimes \left(\tr_E \circ\phi_\calG^\prime \right)
\right)
\left(\psi_\calG\otimes {\bm 0}_E\right)\right]
\nonumber
\\
&
\le 
\min_{\phi_\calG^\prime \in X_\calG}
\max_{\psi_\calG\in\calG}
S\left[
\left(\calI \otimes  U_\phi \right) \left(\psi_\calG\otimes {\bm 0}_E\right)
\|
\left(
\calI \otimes  \phi_\calG^\prime 
\right)
\left(\psi_\calG\otimes {\bm 0}_E\right)\right]
\nonumber
\\
&
\le 
\min_{\phi_\calG^\prime \in X_\calG}
\max_{\psi_\calG^\prime\in\calG}
S\left[
\left(
\calI \otimes  U_\phi  \right)
\left(
\psi_\calG^\prime\right) 
\|
\left(\calI \otimes  \phi_\calG^\prime \right)
\left(\psi_\calG^\prime\right)
\right]
\nonumber
\\
&
=D_\calG \left[U_\phi\right].
\end{align}
The first inequality is from limiting the set of operations $\phi_\calG$ over which the minimization is performed; the second inequality is from relative entropy's monotonically decreasing under a partial trace; and the third inequality is from expanding the set of states over which the maximization is performed.
\end{proof}

\item   \label{Prop:D6} 
Non-increasing under concatenation with a Gaussian channel. $\forall \xi_\calG\in X_\calG^L$, (1) Post-concatenation: 
$
D_\calG\left[\xi_\calG \circ \phi \right]\le D_\calG\left[\phi \right].
$
(2) Pre-concatenation: 
$
D_\calG\left[\phi \circ \xi_\calG \right]\le D_\calG\left[\phi \right].
$

\begin{proof}
(1) $D_\calG\left[ \xi_\calG \circ \phi\right]$
\vspace{-0.5em}
\begin{align}
&=\min_{\phi_\calG\in X_\calG}\max_{\psi_\calG\in\calG}S\left[
\left(\calI \otimes \left( \xi_\calG \circ\phi  \right)\right) \left(\psi_\calG\right)
\|
\left(\calI \otimes \phi_\calG\right) \left(\psi_\calG\right)\right]
\nonumber
\\
&\le \min_{\phi_\calG^\prime \in X_\calG}\max_{\psi_\calG\in\calG}S\left[
\left(\calI \otimes \left( \xi_\calG\circ \phi  \right)\right) \left(\psi_\calG\right)\|
\left(\calI \otimes \left( \xi_\calG\circ \phi_\calG^\prime \right)\right) \left(\psi_\calG\right)\right]
\nonumber
\\
&\le \min_{\phi_\calG^\prime \in X_\calG}\max_{\rho_\calG\in\calG}
S\left[
\left(\calI \otimes  \phi \right) \left(\rho_\calG\right)\|
\left(\calI \otimes  \phi_\calG^\prime \right) \left(\rho_\calG\right)\right]=D_\calG\left[ \phi\right]
.
\end{align}
The first inequality is due to limiting to minimization over $\phi_\calG$ that can be written as $ \xi_\calG \circ \phi_\calG^\prime$; and the second inequality is due to relative entropy's monotonically decreasing under a quantum operation.

(2) $D_\calG\left[ \phi\circ \xi_\calG\right]$
\vspace{-0.5em}
\begin{align}
&=\min_{\phi_\calG\in X_\calG}\max_{\psi_\calG\in\calG}S\left[
\left(\calI \otimes \left( \phi \circ \xi_\calG \right) \right)\left(\psi_\calG\right)
\|
\left(\calI \otimes \phi_\calG\right) \left(\psi_\calG\right)\right]
\nonumber
\\
&\le \min_{\phi_\calG^\prime \in X_\calG}\max_{\psi_\calG\in\calG}S\left[
\calI \otimes \left( \phi \circ \xi_\calG \right)
\left(\psi_\calG\right)
\|
\calI \otimes \left( \phi_\calG^\prime \circ \xi_\calG\right) 
\left(\psi_\calG\right)\right]
\nonumber
\\
&\le \min_{\phi_\calG^\prime \in X_\calG}\max_{\rho_\calG\in\calG}
S\left[
\left(\calI \otimes  \phi \right) \left(\rho_\calG\right)\|
\left(\calI \otimes  \phi_\calG^\prime \right) \left(\rho_\calG\right)\right]=D_\calG\left[ \phi\right]
.
\end{align}
The first inequality is due to limiting to minimization over $\phi_\calG$ that can be written as $\phi_\calG^\prime \circ \xi_\calG $; and the second inequality is due to $\left(\calI\otimes \xi_\calG\right) \left(\psi_\calG\right)\in \calG$.
\end{proof}

\item   \label{Prop:D7} 	
$
D_\calG\left[\phi_1 \otimes \phi_2 \right]\ge \max\left( D_\calG\left[\phi_1\right], D_\calG\left[\phi_2\right]\right).
$

\begin{proof}
$D_\calG\left[\phi_1\otimes \phi_2\right]$
\vspace{-0.5em}
\begin{align} 
&
=\min_{\phi_\calG\in X_\calG}\max_{\rho_\calG\in\calG}
S\left[ 
\left(\calI \otimes \phi_1\otimes \phi_2 \right) \left(\rho_\calG\right)\| 
\left(\calI \otimes \phi_\calG
\right)
\left(\rho_\calG\right)\right]
\nonumber
\\
&\ge
\min_{\phi_\calG\in X_\calG}\max_{\rho_\calG^\prime \in\calG}
S\left[ 
\calI \otimes \phi_1\otimes \phi_2  \left(\rho_\calG^\prime \otimes \sigma\right)\| \calI \otimes \phi_\calG  \left(\rho_\calG^\prime \otimes \sigma\right)\right]
\nonumber
\\
&
\ge
\min_{\phi^\prime_\calG\in X_\calG}\max_{\rho_\calG^\prime\in\calG}
S\left[
\left(\calI \otimes \phi_1 \right) \left(\rho_\calG^\prime \right)
\| 
\left(\calI \otimes \phi_\calG^\prime \right) \left(\rho_\calG^\prime\right) \right]
\nonumber
\\
&
=D_\calG\left[\phi_1\right].
\end{align}
The first inequality is due to limiting to maximization over $\rho_\calG$ that has a product form $\rho_\calG \otimes \sigma$, where $\sigma \in\calG[n_{\phi_2}]$ is fixed. The second inequality is by taking a trace over the input to $\phi_2$ and that $\phi_\calG^\prime \equiv\tr_{2}\circ \phi_\calG$ is a Gaussian channel. 
Similarly, one can prove $D_\calG\left[\phi_1 \otimes \phi_2 \right]\ge D_\calG\left[\phi_2\right]$.
\end{proof}

\end{enumerate}

\section{Covariance matrix and correlations}
\label{App_Cov}

The $4\times 4$ covariance matrix $\Lambda$ of a two-mode (denote them as $A$ and $B$) quantum state $\rho$ can be obtained as follows. Note that $\Lambda=\Lambda^T$. The first diagonal block is given by
\ba
\Lambda\left(1,1\right)&=&2{\rm Re}\braket{a_A^2}_\rho+2\braket{a_A^\dagger a_A}_\rho+1-(2{\rm Re}\braket{a_A}_\rho)^2,
\nonumber
\\
\Lambda\left(2,2\right)&=&-2{\rm Re}\braket{a_A^2}_\rho+2\braket{a_A^\dagger a_A}_\rho+1-(2{\rm Im}\braket{a_A}_\rho)^2,
\nonumber
\\
\Lambda\left(1,2\right)&=& 2{\rm Im}\braket{a_A^2}_\rho-4{\rm Re}\braket{a_A}_\rho {\rm Im}\braket{a_A}_\rho.
\nonumber
\ea
The second diagonal block is given by replacing $A$ with $B$ and $\Lambda\left(i,j\right)$ with $\Lambda\left(i+2,j+2\right)$ in the above equations.

The cross terms are given as follows
\ba
\Lambda\left(1,3\right)&=&2{\rm Re}\left(\braket{a_Aa_B}_\rho+\braket{a_A^\dagger a_B}_\rho\right)-4 {\rm Re}\braket{a_A}_\rho {\rm Re}\braket{a_B}_\rho,
\nonumber
\\
\Lambda\left(2,4\right)&=&2{\rm Re}\left(\braket{a_A^\dagger a_B}_\rho-\braket{a_A a_B}_\rho\right)-4 {\rm Im}\braket{a_A}_\rho {\rm Im}\braket{a_B}_\rho,
\nonumber
\\
\Lambda\left(1,4\right)&=&2{\rm Im}\left(\braket{a_A^\dagger a_B}_\rho+\braket{a_A a_B}_\rho\right)-4 {\rm Re}\braket{a_A}_\rho {\rm Im}\braket{a_B}_\rho,
\nonumber
\\
\Lambda\left(2,3\right)&=&2{\rm Im}\left(\braket{a_A a_B}_\rho-\braket{a_A^\dagger a_B}_\rho\right)-4 {\rm Im}\braket{a_A}_\rho {\rm Re}\braket{a_B}_\rho.
\nonumber
\ea

\section{Mixed unitary channels}
\label{App_mixed_unitary}

The binary phase-shift channel $\phi_{\rm BPS}$ is a probabilistic mixture of Gaussian unitaries. We begin our analysis of it by considering the general case of probabilistic mixing of $K$ Gaussian unitaries $\left\{U_k, 1\le k \le K \right\}$, with probabilities $\left\{p_k, 1\le k \le K \right\}$, i.e.,
\be
\phi_{\rm mix}\left(\rho\right)=\sum_{k=1}^K U_k \rho U_k^\dagger. 
\ee
From Definition~\ref{Def2} and Eq.~(\ref{nG}), with $\rho_{AB}=\calI_{\phi_{\rm mix}} \otimes \phi_{\rm mix}\left(\psi_{AA^\prime}\right)=\sum_{k=1}^K p_k U_k \psi_{AA^\prime} U_k^\dagger$, we have $
0\le S\left(\rho_{AB}\right)\le h\left(\left\{p_k\right\}\right)\equiv-\sum_{k=1}^K p_k\log_2 p_k
$. Let
$
S_\calG^{\rm max}=\max_{\rho_{A^\prime}\in \calG} S\left[\lambda_\calG\left(\rho_{AB} \right)\right]
$.
We have,
\ba
\tdelta_\calG \left[\phi_{\rm mix}\right]&=&\max_{\rho_{A^\prime}\in\calG}S\left[\lambda_\calG\left(\rho_{AB} \right)\right]-S\left(\rho_{AB} \right)
\\
&\in& \left[S_\calG^{\rm max}-h\left(\left\{p_k\right\}\right),S_\calG^{\rm max}\right].
\ea

Because $h\left(\left\{p_k\right\}\right)$ is finite, if one can show that either $S_\calG^{\rm max}$ or $\tdelta_\calG \left[\phi_{\rm mix}\right]$ diverges, then the rate of divergence of $\tdelta_\calG \left[\phi_{\rm mix}\right]$ is the same with $S_\calG^{\rm max}$.

For the case of $\phi_{\rm BPS}$, we have $h\left(\left\{p_k\right\}\right)=1$ and when the output and ancilla have total energy $N_S$,
$
S_\calG^{\rm max}=2 g\left(N_S/2\right)
$
is achieved by input-ancilla in a TMSV.

\flushleft

%\bibliographystyle{myabbrvnat}

%\bibliography{myref}
\newcommand{\cmmnt}[1]{\ignorespaces}

\end{document}